\documentclass[12pt,preprint]{aastex}
\usepackage{graphics, ulem}
\input psfig.sty
\input epsf
\begin{document}
\baselineskip=12pt

\title{INTERSTELLAR TURBULENCE II: IMPLICATIONS AND EFFECTS}
\markboth{\sc Scalo \& Elmegreen}{\sc Interstellar Turbulence II}

\author{John Scalo \affil{Department of Astronomy, University of Texas,
Austin, Texas 78712; e-mail: parrot@astro.as.utexas.edu}}

\author{Bruce G. Elmegreen \affil{IBM Research Division,
Yorktown Heights, New York 10598; email: bge@watson.ibm.com}}


\begin{abstract}
Interstellar turbulence has implications for the dispersal and
mixing of the elements, cloud chemistry, cosmic ray scattering,
and radio wave propagation through the ionized medium. This review
discusses the observations and theory of these effects.
Metallicity fluctuations are summarized, and the theory of
turbulent transport of passive tracers is reviewed.  Modeling
methods, turbulent concentration of dust grains, and the turbulent
washout of radial abundance gradients are discussed.  Interstellar
chemistry is affected by turbulent transport of various species
between environments with different physical properties and by
turbulent heating in shocks, vortical dissipation regions, and
local regions of enhanced ambipolar diffusion. Cosmic rays are
scattered and accelerated in turbulent magnetic waves and shocks,
and they generate turbulence on the scale of their gyroradii.
Radio wave scintillation is an important diagnostic for small
scale turbulence in the ionized medium, giving information about
the power spectrum and amplitude of fluctuations. The theory of
diffraction and refraction is reviewed, as are the main
observations and scintillation regions.
\end{abstract}

\section{INTRODUCTION}

One of the most important developments in the field of
interstellar gas dynamics during the last half-century was the
renewed perception that most processes and structures are strongly
affected by turbulence. This is a paradigm shift unparalleled in
many other fields of astronomy, comparable perhaps to the
discovery of extrasolar planets and cosmological structure at high
redshift.  Interstellar turbulence and its implications were
commonly discussed in the 1950s, but without the range of
observations that are available today, the theory of the
interstellar medium (ISM) drifted toward increasingly detailed
models based largely on a preference for thermal and dynamical
equilibria. Standing apart were two subfields that continued to
include turbulence as the primary driver of all observations:
radio scintillation and cosmic ray transport.  The connections
between these subfields and the prevailing picture of the ISM were
mostly ignored because of the large discrepancy in scales.
Consequently, two other small-scale processes got lost in the
equilibrium paradigm of the large-scale models: mixing of the
elements and molecular chemistry. In this review, we discuss these
four subfields in some detail. All have a well-established
observational base going back several decades, but because of the
complexity of turbulence and the infancy of the relevant theory,
all of them are now in a state of rapid evolution.

Our previous review {\it Interstellar Turbulence I} (this volume),
emphasized observations in the dense neutral ISM and discussed in
detail the various theoretical approaches to this field.

\section{TURBULENCE AND CHEMICAL INHOMOGENEITY}
\label{sect:mixing}

The degree to which turbulence and other transport processes mix
or homogenize the gas in the face of repeated local pollutions by
supernovae and other sources of fresh metals could provide an
important constraint on the hydrodynamics of the ISM. This general
problem is called passive scalar turbulence because the particles
being transported have no effect on the velocity field. A review
of experiments involving incompressible turbulent scalar transport
can be found in Warhaft (2000), who emphasized that passive scalar
probability density functions (pdfs) usually have strong
exponential tails. A review focussing on a particular class of
theoretical approaches has been given by Falkovich, Gawedzki \&
Vergassola (2001), whereas reviews of methods to calculate the
concentration pdf and its moments are in Dopazo (1994) and O'Brien
(1980). Briefer and more specialized treatments can be found in
Shraiman \& Siggia (2000) and Pope (2000, section 12.4).

Here we summarize observations of abundance variations in stars
and gas, and we review recent theoretical work on interstellar
turbulent mixing. We do not discuss star-to-star variations in
globular clusters that may be affected by stellar evolution (Yong
et al. 2003).

\subsection{Metallicity Fluctuations: Observations}

Recent observations of metallicity dispersions in stars of the
same age and in interstellar gas suggest that the ISM is very well
mixed on average, with typical fluctuations about the mean of only
5--20\%. This low value is a challenge to explain considering the
spotty pollution of new elements from supernovae and other
sources. In addition, there are much larger observed fluctuations
in a few places, suggesting the overall distribution function for
abundance has a long, non-Gaussian tail. Both the low dispersion
and the long tail might be explained by interstellar turbulence.

Stars of a similar age have often shown large variations in
metallicity (a term usually referring to iron abundance relative
to the Sun). For example, Edvardsson et al. (1993) and Chen et al.
(2000) found factor-of-four abundance variations in several
elements for F and G main sequence stars. After correcting for
selection effects, Edvardsson et al. estimated that the dispersion
at a given age is about 0.15 to 0.20 dex, i.e., 40--60\%, whereas
Garnett \& Kobulnicky (2000) suggested it is smaller than 0.15 dex
for the same sample. A large unbiased sample of over 550 Hipparcos
stars with homogeneous chromospheric age estimates and photometric
metallicities showed a dispersion at a given age of only 0.13 dex
(Rocha-Pinto et al. 2000). Reddy et al. (2003) studied 27 elements
in 181 F and G stars and found a dispersion at a given age that is
somewhat smaller than in Edvardsson et al. (1993) and Chen et al.
(2000). Feltzing \& Gonzales (2001) used an inhomogeneous sample
of 5828 stars and found a larger scatter in the metallicity at a
given age of 0.2 to 0.3 dex, which exceeds the observational
errors. A major problem is that the ages of the stars used in
these studies have large uncertainties. Considering that the
stellar orbits probably migrated and led to additional abundance
variations from Galactic radial gradients (e.g., Wielen et al.
1996), the stellar results are consistent with a dispersion in the
gas of 20--30\% or less, the upper limit set by uncertainties
inherent in field star samples.

Other evidence for homogeneity among field stars comes from
elemental abundance ratios. Reddy et al. (2003) estimated the
intrinsic scatter in the ratio [X/Fe] for element X by fitting a
Gaussian curve to the histogram of deviations between [X/Fe] and
the linear fit of [X/Fe] versus [Fe/H]. The standard deviations in
the Gaussian for 26 elements in 181 stars varied between 0.03 and
0.10 dex.  These deviations are comparable to the errors, so the
result is an upper limit to the intrinsic scatter. It does not
depend on stellar age, as do the absolute abundances. Reddy et al.
noted that because the elements come in differing proportions from
different sites of stellar nucleosynthesis, the lack of scatter
implies all the ejecta were well-mixed in the gas before the stars
formed.

Abundance fluctuations of 0.05--0.15 dex had also been found in
studies of open clusters (e.g., Friel \& Boesgaard 1992, Twarog et
al. 1997, Carraro et al. 1998), B stars in star-forming regions
(e.g., Cunha \& Lambert 1994) and clusters (Rolleston et al.
1994), the Magellanic clouds (e.g., Olszewski et al. 1991), the
interstellar medium (Meyer et al. 1998), Galactic {\it H II}
regions (Deharveng et al. 2000), and {\it H II} regions in disk
(see review in Henry \& Worthey 1999) and dwarf irregular (e.g.,
Thuan et al. 1995) galaxies.

However, F stars in nearby open clusters (Friel \& Boesgaard 1992
and references therein) show no evidence for star-to-star
abundance variations at the level of detectability.
Cluster-to-cluster variations are also small, including clusters
of all ages, and essentially zero (less than 10--20\%) for the
four clusters younger than $4\times10^8$ years. The latter result
implies that the ISM is mixed, at least to this level, for scales
between $\sim$1 pc (the size of the clumps from which clusters
form) and $\sim$100 pc (the distances between clusters) on a
timescale less than $4\times10^8$ years. Even more stringent
constraints were found from the photometric metallicity study of
76 clusters by Twarog et al. (1997). Comparison of high-precision
photometry with stellar evolution models suggested a metallicity
spread less than $\sim$0.03 dex in the Hyades cluster (Quillen
2002). Paulson, Sneden \& Cochran (2003) derived the same small
[Fe/H] spread, excluding three outlier stars, using spectroscopic
abundance determinations for individual stars. These results imply
fluctuations of less than $\sim$7\% in the gas from which the
Hyades cluster formed.

Interstellar absorption studies suggest equally low dispersions.
Meyer et al. (1998), Moos et al.(2002), and Andre et al. (2003)
found oxygen abundance dispersions over distances of around one
kpc that are only a few percent. A recent study of interstellar
Kr/H by Cartledge, Meyer \& Lauroesch (2003) found a spread about
the mean of only 0.06 dex for lines of sight within the local
Orion spur; this was less than the measurement error.

We conclude that the most recent evidence, based on field stars of
a given age, cluster stars, and the ISM, suggests a metallicity
dispersion that is very small, less than 20 to 30\% and perhaps as
small as a few percent. This small Galactic dispersion is
consistent with the study of 41 {\it H II} regions in M101 by
Kennicutt \& Garnett (1996).

There are notable exceptions, though, including: ({\it a}) the
existence of old super-metal-rich stars (e.g., Feltzing \&
Gonzalez 2001), and a large range in abundance ratios among
metal-poor stars (Fields et al. 2002); ({\it b}) the lack of an
age-metallicity relation for five nearby clusters, which led
Boesgaard (1989) to conclude that local abundance fluctuations of
$\sim$0.2--0.3 dex must persist for at least $10^8$--$10^9$ years
(but see Friel \& Boesgaard 1992 for a revised result); ({\it c})
the factor of 2 to 3 fluctuations in [O/H] among subgroups in
Orion (Cunha \& Lambert 1994), which, however, may be a result of
self-pollution; ({\it d}) the study of B stars in several clusters
at galactocentric distance $\sim$13 kpc that indicates one cluster
has [Fe/H] larger than the rest by a factor of $\sim$5 (Rolleston,
Dufton \& Fitzsimmons 1994), ({\it e}) two stars in the Carina
nebula (Andre et al. 2003) that are separated by only a few
hundred parsecs but have line-of-sight oxygen abundances different
by about 40\%, and ({\it f}) outlier stars in the cluster studies
discussed earlier.

These examples suggest a metallicity pdf with a small variance and
possibly large higher moments, i.e., a fat tail at large
metallicities. Such a pdf is similar to that found for the pdf of
passive scalars in incompressible turbulence (Warhaft 2000). The
analogous pdf for a compressible, magnetic, and self-gravitating
ISM is unknown.

\subsection{How Large Should Metallicity Fluctuations Be?}
\label{sect:howlarge}

Theoretical investigations of abundance fluctuations have until
recently nearly all been based on order of magnitude arguments
involving characteristic spatial scales and timescales for various
processes. Two of the more thorough discussions are in Roy \&
Kunth (1995) and Tenorio-Tagle (1996). Regarding spatial scales,
Elmegreen (1998) pointed out that for a hierarchically structured
ISM with star formation operating on a local dynamical time, the
size of an O and B star association is usually larger than the
supernovae remnants it produces. Unless there is some immediate
mixing, as in hot ionized or evaporating flows, the new elements
should be spotty and variable for subsequent generations of stars
(self-enrichment). The point about timescales was highlighted by
Reeves (1972): The mean time for a gas element to get
reincorporated into a star is shorter than the time for its
dispersal by supernova mixing, so abundance fluctuations should be
large. Both results contradict the apparent uniformity of
abundances inside and between clusters and in the general ISM.

Estimates for mixing times have changed since Reeves' (1972)
paper. Edmunds (1975) found that the timescale for mixing of gas
by turbulence and galactic shear is relatively short compared with
the star formation time, provided the sources of metals do not
cluster together making the dispersion more difficult (Kaufman
1975). Bateman \& Larson (1993) also derived a short mixing time,
considering diffusion by cloud motions, as did Roy \& Kunth
(1995), who applied the diffusion approximation to galactic shear,
radial flows, turbulent diffusion, supernovae, and gas
instabilities. Similarly short timescales can be derived using
structure functions by considering the separation of initially
nearby Lagrangian fluid particles (Frisch 1995). This latter
procedure is often used to derive the Richardson 4/3 law of
superdiffusive incompressible turbulence. The implication of these
short mixing times is that the ISM should be homogeneous.

The expectation of large spatial fluctuations may be viewed in a
different way too, by considering the number of overlapping
regions of contamination. Edmunds (1975) showed that $n$
supernovae in a local volume should lead to inhomogeneities of
order $n^{-1/2}$ if there is no subsequent mixing. Adopting an
average of $n\sim5\times10^4$, he suggested that fluctuations
would be relatively small. More detailed models for overlapping
contaminations were studied by Roy \& Kunth (1995), Argast et al.
(2000) and Oey (2000). Oey (2003) used turbulent dispersal from
Bateman \& Larson (1993) to enlarge the effective sizes and reduce
the average metallicities in the overlapping regions.  Low
metallicity stars and galaxies should have large metallicity
dispersions because the numbers of contaminating events are small
(Audouze \& Silk 1995).

A problem with most of the timescale arguments is that they apply
to turbulent transport and not homogenization. Moving clouds
around or shredding them with fluid instabilities does not
homogenize the gas at the atomic level or change the frequency
distribution of abundance concentrations; viscosity and molecular
diffusion are required for that.  The same is true for
incompressible turbulence imagined as a cascade of vortices.
Unless the advection process creates steep gradients on small
scales, molecular diffusion cannot occur in a reasonable time. The
situation changes when there are large-scale gradients in the mean
concentration; then advection from distant regions can change the
concentration distribution locally (see Equation~\ref{eq:Zave}
below). Even so, homogenization requires molecular diffusion.
Thus, a small transport time in the ISM does not imply the gas
will homogenize enough to remove abundance differences before it
forms stars and clusters. The size and placement of each
protostellar core within the complex spatial pattern of abundance
fluctuations will determine its metallicity.

\begin{figure}
\plotone{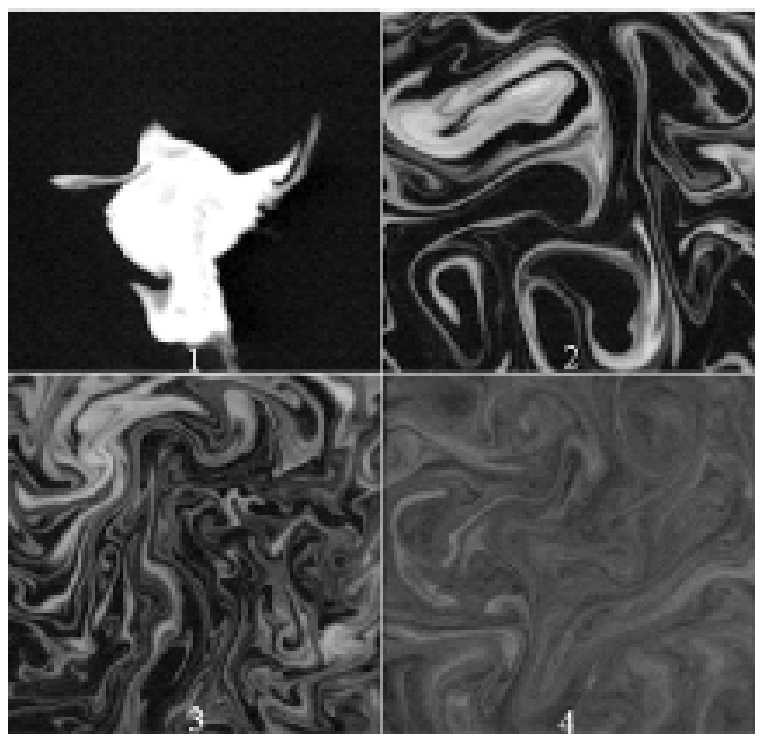} 
\caption{This sequence of four
time steps shows a blob of dye released into an incompressible
two-dimensional laboratory turbulent fluid that is forced by a
chaotic velocity field with a single large scale.  The evolution
of this concentration field illustrates some of the basic features
that can be expected for passive scalars like newly produced
elements or dust grains released into the ISM: Nonlinear advection
transports concentration away from the source but concentrates it
locally through stretching and folding into thin regions of large
velocity gradients where molecular diffusion eventually results in
true mixing.  From Jullien et al. (2000).{\it (for higher resolution,
see paperII-f1.gif)}}
\label{fig:mix}\end{figure}

Releasing metals into the interstellar turbulent gas is similar in
basic respects to dropping a blob of ink into a turbulent fluid.
Figure \ref{fig:mix} shows such a release into a two-dimensional
(2D) incompressible forced chaotic velocity field (Jullien et al.
2000).  Advection spreads the scalar out in space, while
concentrating it locally through stretching and folding, shearing
it into thin ribbons until the spatial gradients are large enough
that molecular diffusion becomes significant in the third frame.
By the final frame, diffusion has smeared out most of the striated
structures.

How does turbulence deliver gradients to small enough scales for
molecular diffusion to cause homogenization in a reasonable time?
de Avillez \& Mac Low (2002) found a surprisingly long timescale
for the decay of the variance in a passive scalar field that had
an initial checkerboard pattern and was mixed by supernovae. Their
time scaled to a galaxy (at least a few times $10^8$ years) is
larger than previous estimates and still is a lower limit because
they did not include sources of new metals. de Avillez \& Mac Low
also found that the mixing time is independent of the mixing
length, unlike mixing modeled as diffusion, and that the mixing
time decreases with increasing supernova rate as a result of the
increased velocities.

There may be ways other than turbulence to get gradients on small
enough scales for molecular diffusion to homogenize. In
Tenorio-Tagle's (1996) ``fountain with a spray'' model, metal-rich
droplets falling onto the disk are subject to Rayleigh-Taylor
instabilities that reduce their size even further; diffusion will
be very effective in such a situation.

\subsection{Methods of Analysis}

There are many methods for studying passive scalar turbulent
transport. Here we describe two approaches: prescribed artificial
velocity fields and closure methods for studying restricted
properties of turbulent transport.

\subsubsection{Artificial stochastic velocity fields}

Most of what is known about passive scalar turbulence comes from
models using an idealized velocity field that is not a rigorous
solution to the hydrodynamics equations (see review by Falkovich
et al. 2001). An example would be a time-uncorrelated velocity
field with Gaussian statistics, a prescribed spatial correlation
function, and a constraint of incompressibility. This is called a
Kraichnan velocity field (e.g., Shraiman \& Siggia 2000). The
advantages of such a field are that the equations for the
concentration field and its statistical properties can in some
cases be solved analytically without a closure assumption, and the
generic properties of the result might be valid for any velocity
field that has the same variance and conservation constraints.
However, there is a large danger in assuming a delta function for
the time correlation function.

Large intermittency in the velocity gradient field means that a
large fraction of the fluid contains large gradients compared with
a Gaussian. In incompressible turbulence (and the ISM---see {\it
Interstellar Turbulence I}), this fraction gets larger on smaller
scales. This implies that intermittency should homogenize the gas
faster than purely Gaussian motions, because molecular diffusion
depends on large velocity gradients at small scales.

Decamp \& Le Bourlot (2002) used another model velocity field that
was obtained from a wavelet reconstruction with statistical
properties similar to the ISM velocity field, including motions
correlated in space and time. They solved the continuity equations
and showed that the correlated velocity field has a faster
dispersion and a larger standard deviation for passive scalars (by
factors of 2 to 4) than an uncorrelated Gaussian velocity field.
The spatial distributions for species undergoing chemical
reactions were also different, suggesting a mechanism for
molecular segregation in dense clouds (Section \ref{sect:chem}).
Further work on turbulent mixing using synthetic velocity fields
is in Elliott \& Majda (1996), Fung \& Vassilicos (1998), and
Boffetta et al. (1999).

\subsubsection{Closure methods: PDF and moment equations}

For some purposes, it may be sufficient to understand how the
mean, variance, and skewness of a metallicity distribution depend
on the source terms and large-scale gradients. This is the moment
approach, and it has the advantage that some of its results can be
obtained analytically. The moment approach might explain, for
example, why there is an inverse square root scaling on the
supernova rate for the decay timescale of the abundance variance
(de Avillez \& Mac Low 2002). The evolution equation for the
one-point probability distribution of metallicities was derived
rigorously for incompressible turbulence by Dopazo, Valino \&
Fueyo (1997), who also derived the moment equations for
incompressible flow (see also O'Brien 1980; Chen \& Kollman 1994;
Dopazo 1994; Pope 1994, 2000).

Perhaps the simplest question that can be asked about scalar
turbulence concerns the way in which the average relative
concentration $<Z>$ of a local patch of contaminant (relative to
hydrogen, for example) will spread in space as a function of time.
The usual place to start is to examine the mean square distance,
$<x^2(t)>$, traveled by a marked Lagrangian fluid particle. This
distance can be related to an integral over the Lagrangian
correlation function, assuming only statistically stationary and
isotropic turbulence.  Without knowing the correlation function,
however, this relation is of little use.  An approximation is to
use the fact that the correlation is unity at time zero and
approaches zero for large separations.  Then for times much
smaller than the correlation time $T_L = \int_0^\infty
R(\tau)d\tau$, $x_{\rm rms} = v_{\rm rms}t$ whereas at $t>>T_L$,
$x_{\rm rms} = 2^{1/2} v_{\rm rms}\left(T_Lt\right)^{1/2}$.  Here,
$R(\tau)$ is the two-time correlation function at a given point in
the flow for lag $\tau$; i.e., $R(\tau)= <u\left(t\right)
u\left(t+\tau\right)>/\sigma^2$ for flow speed $u$ and the average
$<>$ is over all pairs of points separated by $\tau$.
Normalization to the variance, $\sigma^2=<u\left(t\right)
u\left(t\right)>$, makes $R(\tau)$ dimensionless with values
between $-1$ and $1$. In this formulation, the particle behaves
ballistically at short times and has an uncorrelated random walk
at large times. These results are usually referred to as Taylor's
theorem (Taylor 1921; see McComb 1990; Pope 2000). The result is
important for illustrating the nature of turbulence as a
correlated random walk and the analogy between turbulent and
molecular diffusion, where in the latter case $T_L$ would be the
mean collision time.

Klessen \& Lin (2003) recently showed that Taylor's theorem is
correct in 3D-compressible turbulence. McComb (1990), Piterbarg \&
Ostrovskii (1997) and others have pointed out that Taylor's
theorem does not solve the problem of turbulent dispersal and
transport, even for the mean field because what is required is
information in the lab frame, not following the particle. Klessen
\& Lin (2003) suggest that the result can be used to develop a
phenomenological mixing-length or diffusion model for turbulent
transport in the ISM in which the integral correlation time is
related to the mean time between shock passages.

Another fundamental relationship is Richardson's law (Richardson
1926) for the rate at which pairs of points separate in space in
the inertial range of turbulence. One form of this law states that
the time dependence of the rms separation is $t^{3/2}$ (Falkovich
et al. 2001, Boffetta \& Sokolov 2002, Nicolleau \& Vassilicos
2003). Work on pair dispersion using synthetic velocity fields has
resulted in at least two important insights: ({\it a}) The
dynamics of particle pairs is sensitive to initial conditions but
their rms separation increases algebraically, not exponentially
(unlike low-dimensional chaotic systems). ({\it b}) Particle pairs
travel together for long times and then separate explosively when
they encounter straining regions around hyperbolic points in the
flow (e.g., Nicolleau \& Vassilicos 2003).

A traditional approach that could be applied to the spreading of
newly produced elements from sources in the ISM is to ignore the
detailed dynamics and use an equation for the evolution of the
mean concentration field alone.  In general, the concentration $Z$
of an element evolves in a given velocity field ${\bf u}$ as
\begin{equation}
\partial Z\left({\bf x},t\right)/\partial t + {\bf u}\left({\bf
x},t\right) \cdot {\bf \nabla} Z\left({\bf x},t\right) =
 {\bf \nabla}\cdot k\left({\bf x},t\right)\nabla Z\left({\bf
x},t\right) + S\left({\bf x},t\right) , \label{eq:Z}
\end{equation}
where $k$ is the mass diffusivity and $S$ is the rate at which
sources inject the element. The second term on the left represents
the effect of the turbulent velocity field in advecting the
concentration field. For the interstellar medium this velocity
field will consist of complex compressions and vortical motions.
The first term on the right represents molecular diffusion, which
operates only on small scales in regions where the gradient of the
concentration field is large. This equation illustrates how
turbulent advection of a passive scalar spreads it out in space
while increasing the local gradients through stretching,
compressing, and folding. Molecular diffusion then operates where
the gradients are high.

Timescales for molecular diffusion have been discussed recently by
Oey (2003) for a uniform gas. The result is sensitive to
temperature, so most of the homogenization of elements in the ISM
may occur in the hot regions, like superbubbles (Silich et al.
2001), as well as those with the steepest velocity gradients.

To obtain the time evolution of the average concentration, one
might choose to ignore all the details of the turbulence and
spatially average Equation \ref{eq:Z}. Ignoring also the molecular
diffusion and source terms, we get
\begin{equation}
\partial <Z>/\partial t + <{\bf U}>\cdot\nabla<Z>
= -<{\bf u}\cdot{\bf \nabla}z> = -{\bf \nabla}\cdot<{\bf
u}z>+<z{\bf\nabla}\cdot {\bf u}> \label{eq:Zave}\end{equation} for
mean concentration and velocity $<Z>$ and $<{\bf U}>$ and
fluctuating parts $z$ and ${\bf u}$. The last term is zero for
incompressible turbulence, leaving only $-{\bf \nabla}\cdot<{\bf
u}z>$, which contains the unknown ensemble average of the scalar
flux, $<{\bf u}z>$, representing interactions between the
fluctuating parts of the velocity and concentration fields.  This
is a classic closure problem. We could derive a differential
equation for the scalar flux, but it would contain more unknown
correlations, including triple correlations, and further
manipulation only leads to higher and higher order unknowns.

In the incompressible case, the simplest treatment is to imagine
that the turbulent transport behaves diffusively, like molecules
in a gas. Then the scalar flux is proportional to the mean
concentration gradient and given by
\begin{equation}
<{\bf u}z>=-K_T{\bf\nabla}<Z>,\label{eq:flux}
\end{equation}
where $K_T$ is a scalar diffusivity given by some timescale times
the square of some velocity.  Often these are taken to be the
correlation time and the rms velocity, although a variety of other
forms have been used to fit specific experimental results
(Piterbarg \& Ostrovskii 1997).  Using Equation~\ref{eq:flux},
Equation~\ref{eq:Zave} becomes the usual diffusion equation, and
the patch of concentration, if initially a point source, would
spread according to a Gaussian whose width grows $r\sim t^{1/2}$.

In fact, turbulent transport cannot be described by classical
diffusion because triple correlations are as important as
quadratic correlations, or, more physically, turbulent motions are
not random independent steps but correlated bulk transports over a
wide range of scales.

There have been many attempts, all unsuccessful to various
degrees, to develop closure techniques for passive scalar
turbulence.  Many of these are described in reviews by O'Brien
(1980) and Dopazo (1994).  Perhaps the most popular in the physics
community is the mapping closure suggested by Chen, Chen \&
Kraichnan (1989). A more complicated closure method is based on
the Eddy-Damped Quasinormal Markovian approximation (Lesieur
1990). An application of this closure to the passive scalar
problem with reference to earlier work is in Herr, Wang \& Collins
(1996).

A promising approach that is not too complex yet captures the
importance of the triple correlations was proposed by Blackman \&
Field (2002) in connection with turbulent dynamo theory and then
applied to the scalar turbulence problem (Blackman \& Field 2003).
The idea is to model the time derivative of the scalar flux as a
normal diffusion term with an eddy viscosity plus a damping term
$<{\bf u}z>/\tau$,  where $\tau$  is a relaxation time.  This
approach is similar to a closure method used in terrestrial
atmospheric turbulent transport (Lewellen 1977). When this is
substituted in the equation for $<Z>$, a little manipulation
results in a damped wave equation for the mean concentration.
Brandenburg, Kapyla \& Mohammed (2004) did extensive comparisons
with numerical simulations of passive scalars in mildly
compressible turbulence (rms Mach number $\sim$0.2) and found good
agreement with this model for the time evolution of the spreading
size and the non-Gaussianity (kurtosis) of $<Z>$.

The moment approach can derive the variance of the abundance
distribution directly from Equation~\ref{eq:Z} by writing the
velocity, abundance, and source fields as the sum of mean and
fluctuating parts (e.g., $Z = <Z> + z$), multiplying the resulting
equation for $\partial z/\partial t$ by $z$, and then ensemble
averaging to obtain
\begin{equation}
\onehalf d<z^2>/dt  = - \onehalf <z{\bf u}> \cdot{\bf \nabla}<Z> -
\onehalf<{\bf U}>\cdot{\bf \nabla} <z^2> - \onehalf<{\bf
u}\cdot\nabla z^2> + <sz> +\; {\rm
diffusion},\label{eq:z2}\end{equation} where $s$ is the
fluctuating part of the source term ($<s> = 0$). The first term on
the right shows how the scalar flux ($<z{\bf u}>$) interacts with
any large-scale gradient in the abundance, increasing variance if
$<z{\bf u}>$ is proportional to a mean gradient (as in
Equation~\ref{eq:flux}).  This is analogous to how stellar orbit
diffusion in the presence of a radial gradient produces
fluctuations in the stellar abundances. The second term indicates
that any gradient in the mean square fluctuations will be advected
with the mean flow (e.g., radial flows or infall).

The third term in Equation~\ref{eq:z2} shows how turbulence can
affect a concentration variance even with no gradient in $<Z>$,
zero mean velocity, and without diffusion. Writing this term as
$<\nabla\cdot\left({\bf u}z^2\right)>-<\left(\nabla\cdot
u\right)z^2>$ shows that it consists of the mean divergence of the
flux of scalar variance, as might be expected, minus a term that
is only nonzero in the presence of compressibility.
Differentiation and ensemble averaging commute, so the first term
of these two is zero for a homogeneous fluid because ensemble
averages do not depend on spatial position. The fourth term
represents the net production of scalar variance by the
fluctuating nucleosynthetic sources. ``Diffusion'' in this context
is shorthand for the fairly complex set of second-order
derivatives involving mass diffusion at the molecular level.

As a simple illustration, assume there are no large scale flows
and the abundance is linear in galactocentric distance with
constant of proportionality $G$. For a general gradient term, we
would have the triple correlation whose closure is discussed by
Blackman \& Field (2003) and Brandenburg, Kapyla \& Mohammed
(2004). Then we can also derive an equation for the correlation
$<z{\bf u}>$ by multiplying Equation~\ref{eq:Z} by ${\bf u}$ and
ensemble averaging. The result is, for $<{\bf U}>=0$,
\begin{equation}\partial<zu_i>/\partial t = - <u_i u_j>G -
<z \left(\partial p/\partial x_i\right) /\rho > + <su_i> + \;{\rm
diffusion}.\label{eq:zu}\end{equation} The first term on the right,
which is $<u^2>G$ for an isotropic velocity field, represents the
production of correlations by interaction of the velocity
fluctuations with the mean metallicity gradient.  We can assume
that we are given some steady-state isotropic turbulent velocity
field so that $<u^2>$ is constant.  The second term on the right,
often called pressure scrambling, is a serious closure problem
for compressible turbulence. The source term is important if the
metals come from young stars whose formation is associated with
the turbulence itself, but for Type I supernovae or any other
source that has time to decorrelate from the flow, $<s{\bf u}>=0$.
This indicates the importance of using a consistent model for the
formation of metal-producers from the turbulent field. Equations
\ref{eq:z2} and \ref{eq:zu} can now be solved to give the
evolution of the standard deviation of the abundance distribution
for a particular velocity and source field.

If we neglect pressure scrambling in Equation~\ref{eq:zu} and the
compressibility term in Equation~\ref{eq:z2}, and also replace the
triple correlation term with a simple gradient-diffusion closure,
then the result would show that the concentration fluctuation
variance increases $\propto<u^2>G^2t^2$, even in the absence of
sources. The turbulence is transporting high-z gas into low-z gas
(and vice versa) due to the existence of the gradient. Obviously,
after some time the large fluctuations imply steep gradients at
very small scales, and so the microscopic diffusion term will
cause the variance to level off. In addition, the gradient itself
will be washed out by the turbulence unless there is some
mechanism continually producing it. Without a sustained mean
gradient, such equations could be used to understand how the
variance in elemental fluctuations should decrease with time under
the action of only advection and diffusion, giving a simplified
model to understand the numerical decay results of de Avillez \&
Mac Low (2002). With some assumed closure for the source terms, or
with prescribed velocity field statistics, the solution of
Equation~\ref{eq:z2} would show how the abundance variance depends
on the source rate.

\subsection{How Do The Metals Enter the Flow?}
\label{sect:enterflow}

The source terms in the transport equations depend on how the
metals in expanding supernovae and superbubbles are released into
the turbulent background. This is not a trivial problem, as
discussed by Tenorio-Tagle (1996). Simulations of superbubbles
(Korpi et al. 1999) and supernova remnants (Balsara et al. 2001)
expanding into a turbulent ISM may provide guidance. Some
metal-producing outflows should be jet-like, leading to the
release of metals by shear flow along the walls of a chimney.
Begelman \& Fabian (1990) and Slavin, Shull \& Begelman (1993)
proposed that such turbulent layers could mix hot and warm gas,
but they could just as well mix freshly-produced metals.
Tenorio-Tagle (1996) suggested that superbubbles blow out of the
disk and rain down metal rich droplets that break apart by
Rayleigh-Taylor instabilities. This process could disperse metals
over several kpc in the disk with no assistance from disk
turbulence.

These processes are also relevant for the redistribution of
elements ejected by galactic blowout into the intergalactic medium
(IGM). Several papers (e.g., Ferrara et al. 2000, Aguirre et al.
2001 and references therein) have noticed that it is difficult to
mix metals on scales comparable to the mean distance between
galaxies, requiring an additional (unknown) mechanism. Turbulence
in the IGM could provide such a mechanism.

\subsection{Turbulent Concentration of Dust Grains}

The advection of the grains by turbulent gas can lead to spatial
clustering relative to the gas, and to a spatially inhomogeneous
dust-to-gas ratio.  The degree of segregation depends on the grain
size.  Local variations in grain-grain velocities affect grain
growth and shattering. Early estimates for the efficiency of this
process were given by Cameron (1973), Burke \& Silk (1976), Scalo
(1977), and Draine (1985), but the conception of how grain
advection occurs differs greatly in these papers. More rigorous
derivations, aimed at grain collisions in protostellar disks but
also applicable to the ISM, were given by V\"olk et al. (1980) and
Markiewicz, Mizuno \& V\"olk (1991).

Recent work has been directed toward trapping grains in
protoplanetary disks by anticyclonic vortices (e.g., Bracco et al.
1999), which can persist for many turnover times. Large grain
concentrations can result, accelerating the formation of
planetesimals. Grains might also become trapped in the general ISM
because of drag forces, as proposed by Falgarone \& Puget (1995).

A review of experiments and simulations on turbulent gas-grain
segregation was presented by Fessler, Kulick \& Eaton (1994). More
recent theoretical investigations (e.g., Sigurgeirsson \& Stuart
2002, Lopez \& Puglisi 2003) have demonstrated several interesting
properties of this clustering. All of this work relies on
synthetic velocity fields because inertial particle advection in
real turbulent flows would require resolution much finer than the
dissipation scale.

Nevertheless, the essential physics of particle drag is becoming
more clear. The particle motion is dissipative even if the
advecting flow is incompressible, and hence, dissipationless
(Balkovsky, Falkovich \& Fouxon 2001). This leads to simplified
treatments in which the Stokes drag effect is modeled as a small
compressible component to the velocity field of the inertial
particles.  A detailed study of inertial particle dynamics in
incompressible flows (Bec et al. 2003) shows that particles
smaller than a critical size form fractal clusters whereas larger
sizes fill space with a nonuniform density. An extension to highly
compressible flows is  desirable.

The motion of charged dust grains in anisotropic
magnetohydrodynamic (MHD) turbulence was considered by Yan \&
Lazarian (2002), who made a distinction between the directions
parallel and perpendicular to the field. Grains with a gyration
time around the field that is shorter than the gas drag time will
follow the turbulent field fluctuations in the perpendicular
direction.  Yan \& Lazarian (2003b) calculated the acceleration of
charged grains that resonate with MHD fluctuations.

\subsection{Turbulent Washout of Radial Abundance Gradients}

Another reason to study stirring by interstellar turbulence is to
understand the origin of galactic radial abundance gradients
(Carraro et al. 1998, Hou et al. 2000, Rolleston et al. 2000).
Models typically account for these gradients using a radially
dependent timescale for infall and star formation (e.g., Chiappini
et al. 2001). A problem is that turbulence should reduce
large-scale gradients, so studies of this turbulent washout could
give an estimate of the timescale on which the gradient must be
replenished. Order of magnitude estimates along the lines of those
mentioned in Section \ref{sect:howlarge} indicate replenishment
timescales larger than characteristic infall times, suggesting
that contributions from a radially dependent stellar initial mass
function (IMF) or other effects are not required.

A peculiar abundance gradient has been observed in the outer
Galaxy, where clusters (Twarog et al. 1997) and Cepheids (Caputo
et al. 2001, Andrievsky et al. 2002, Luck et al. 2003) suggest a
sudden decrease in metal abundance by about 0.2 dex setting in at
10 kpc. This discontinuity appears over a radial distance less
than 1 kpc. It is hard to see how such a feature could survive
turbulent washout for more than  $10^8$ years using timescale
arguments of the type summarized earlier.  Either we are missing
something fundamental about turbulent transport (e.g., maybe
differential rotation or magnetism suppress turbulent motions in
the radial direction), or this metallicity near-discontinuity does
not really exist.

\section{EFFECTS OF TURBULENCE ON INTERSTELLAR CHEMISTRY}
\label{sect:chem}

The coupling of interstellar chemistry to the dynamics of
interstellar gas has focused for a long time on collapsing or
slowly contracting clouds (e.g., Gerola \& Glassgold 1978; Prasad,
Heere \& Tarafdar 1991; Shematovich et al. 2003) and single shocks
or MHD waves (e.g., Hollenbach \& McKee 1979, 1989; Draine \& Katz
1986; Charnley 1998; Flower \& Pineau des For\^ets 1998). The
collapse models explained why early-time solutions for static
chemistry often agreed better with observations than late-time
solutions (e.g., Prasad et al. 1991; Nejad \& Wagenblast 1999; but
see Turner et al. 2000). The result that nondynamical
time-dependent abundances agreed with observations for different
ages is a strong motivation for the influence of turbulence
because turbulent timescales are often smaller than the chemical
equilibrium times. Turbulent transport can prevent the chemistry
from attaining a steady state (e.g., Phillips \& Huggins 1981;
Xie, Allen \& Langer 1995; Willacy, Langer \& Allen 2002).

Turbulence can affect ISM chemistry in three ways: ({\it a}) by
continually transporting material between regions with different
physical conditions, like ambient ultraviolet (UV) radiation flux,
temperature, and density, ({\it b}) by creating localized heating
where temperature-sensitive reactions, especially those involving
endothermic reactions, will be enhanced, and ({\it c}) by
magnetically forcing ions to move faster than the thermal speed
where they can enhance the temperature-sensitive ion-neutral
reactions. Progress in this field is difficult because of
limitations in the dynamical models and because of the large
number of nondynamical chemical effects that may occur, like
neutral-neutral reactions at low temperature, variable C and O
depletion, three-body channels for dissociative recombination, the
presence of cosmic-ray-induced UV photons, chemical phase
transitions, and many more, as reviewed by Herbst (1999),
Hollenbach \& Tielens (1999), van Dishoeck \& Hogerheijde (1999),
and Langer et al. (2000). For more recent work on chemistry
covering a variety of applications not necessarily related to
turbulence, see Roberts \& Herbst (2002), Doty et al. (2002),
Stantcheva \& Herbst (2003), Shematovich et al. (2003), and
Charnley \& Markwick (2003).

\subsection{Turbulent Transport of Chemical Species}

The classic observation that implied turbulent transport was the
large amount of carbon in the gas phase of dense molecular cores,
considering the short freeze-out time on dust. Boland \& deJong
(1982) proposed that turbulent circulation could explain this
discrepancy by bringing grains to the outer layers for UV
desorption.  Modern observations (Kramer et al. 1999) show more
depletion, however, and photodissociation region (PDR) models
suggest carbon comes from clump surfaces at a range of depths
(Howe et al. 2000).

Turbulent transport has other applications in modern chemical
models. Pijpers (1997) considered a diffusive model of turbulent
grain transport and obtained a long migration time. Ragot (1998)
pointed out that turbulent transport is probably neither ballistic
nor diffusive, and used a model based on Levy flights to return to
the smaller timescales. This transport effect is important because
the degree to which CO is released from grains into the gas can
affect the rest of the chemistry (CO deactivates much chemistry by
depleting the H$_3^+$ ion; Nejad \& Wagenblast 1999).

A detailed study of turbulent transport including 87 species and
1100 reactions was given by Xie et al. (1995), who assumed that
radial flux is proportional to the spatial gradient of a species
and used a mixing length model for turbulent diffusion. They found
that most carbon-bearing species and several other important
molecules were strongly affected by turbulence. The model was
extended recently by Willacy, Langer \& Allen (2002), who showed
how the {\it H I}/H$_2$ ratio could provide a sensitive signature
of turbulent transport. This occurs because H$_2$ forms when H
reacts on dust grains. The formation rate depends on the density
of atomic H and the total dust cross section, whereas the
destruction rate depends on the UV field. If turbulence cycles
material between regions with different UV optical depths on a
timescale comparable to the formation time, then H will penetrate
deeper into the cloud and H$_2$ will be closer to the surface.
Because many other simple molecules are formed from H$_2$-based
intermediaries (especially H$_3^+$), turbulent transport could
control the chemistry for many species.

Timescales for reaching chemical equilibrium depend on density,
temperature, and ionization fraction, and are typically
$10^5$--$10^7$ years (van Dishoeck \& Hogerheijde 1999). For the
atomic part of a photodissociation region, the longest part of the
cycle C$^+\rightarrow$ CH$_2^+\rightarrow$ CO$\rightarrow$ C
$\rightarrow $C$^+$ is the formation time of CH$_2^+$ through
radiative association (Tielens \& Hollenbach 1985) with a
timescale $\sim$10$^6n_2^{-1}$ years for $n_2=$ particle density
in units of $10^2$ cm$^{-3}$. For turbulent transport, the fastest
possible timescale is ballistic, $L/v$, for turbulent speed $v$
and length $L$. Using the ISM scaling relations for illustration
(see {\it Interstellar Turbulence I}), this minimum time is
$10^6/n_2^{1/2}$ years. The chemical and transport timescales are
comparable at $10^2$ cm$^{-3}$, but the transport time becomes
larger than the chemical time at higher density and then
turbulence would seem to have little effect. Because cosmic ray
ionization drives the chemistry, and each ionized H$_2$ molecule
eventually forms CO or some other important molecule, the chemical
timescale in the cores of dense clouds is the ratio of the
relative abundance of CO, $\sim$10$^{-4}$, to the cosmic ray
ionization rate, $\sim$4$\times10^{-17}$ s$^{-1}$; this gives
$\sim$10$^5$ years (D.J. Hollenbach, private communication).

\subsection{Turbulent Heating}

Another way in which turbulence may affect chemistry is through
localized heating in shocks (Flower \& Pineau des For${\hat {\rm
e}}$ts 1998), vortices (Joulain et al. 1998), ambipolar diffusion
(Padoan, Zweibel \& Nordlund 2000), and magnetic reconnection
(Lazarian \& Vishniac 1999). There are many reactions that seem to
require higher than normal temperatures.

For diffuse clouds, the large abundance of CH$^+$ and molecules
like HCO$^+$ that can form from CH$^+$, and the large abundance of
OH (Gredel 1997, Lucas \& Liszt 1997), all suggest that high
temperature ($\sim$10$^3$ K) reactions take place, as in shock
fronts (Elitzur \& Watson 1980, Draine \& Katz 1986).  However,
the correlation between CH$^+$ and the lower-temperature molecule
C$_2$ (Gredel 1999), and the similar radial velocities of CH$^+$
and CH rule out CH$^+$ production in single large shocks or cloud
surfaces. Instead, CH$^+$ could be made in numerous unresolved
shocks that blend on a line of sight with cooler gas at the same
average velocity (Gredel, Pineau des For${\hat {\rm e}}$ts \&
Federman 2002). Part of the appeal of turbulence is its great
range of scales that can yield such an effect.

Temperature fluctuations of several hundred degrees could play a
major role in the chemistry of many species because of the inverse
temperature dependence of radiative association and the range of
activation energies for collisional dissociation, neutral-neutral
reactions, and some ion-molecule reactions. Neutral-neutral
reactions may be important even in dark clouds (Bettens et al.
1995).  Prasad \& Huntress (1980) long ago pointed out the
importance of temperature variations in radiative association
reactions and some ion-molecule reactions. Also, in dense shielded
clouds, warm regions drive most of the O, OH, and O$_2$ to H$_2$O
through reactions like O + H$_2\rightarrow$ OH$+$H and
OH+H$_2\rightarrow$H$_2$O+H (e.g., Charnley 1998), not only
enhancing the H$_2$O abundance, but also preventing the formation
of molecules like SO$_2$, which rely on the oxygen species for
their formation. The CN abundance should be enhanced because its
destruction by reaction with O and O$_2$ is suppressed; this can
lead to enhanced production of HC$_3$N through reaction of CN with
C$_2$H$_2$.

There have been several attempts to model the effects of local
turbulent heating on chemistry. Black \& van Dishoeck (1991)
pointed out that if turbulence extends down to the collisional
mean free path, then the tail of the particle velocity
distribution could be enhanced. Reaction rates are averages over
the particle relative velocity distribution, so energy-dependent
cross-sections could have greatly enhanced rates.  Spaans (1996)
noted that the distribution of velocity differences in turbulent
flows at small scales has a large tail and suggested this could
result in non-Maxwellian relative speeds between molecules,
especially those involved with CH$^+$ production. This assumption
is questionable because the reaction scales are the mean free
paths, of order 1AU$/n_2$, and these are smaller than the viscous
dissipation scale of incompressible turbulence. The validity of
the fluid approximation for neutrals is also questionable at these
scales (see {\it Interstellar Turbulence I}),
although MHD turbulence in the ionic component
could enhance the ion-neutral collision rate.

Another approach to the problem was presented in a series of
papers by Falgarone and coworkers.  Falgarone \& Puget (1995),
Falgarone et al. (1995), and Joulain et al. (1998) considered the
diffuse cloud abundances of CH$^+$ and several other molecules
such as OH and HCO$^+$.  They focused on the solenoidal part of
the velocity field in which molecular viscosity on 10 AU scales
produces numerous local hot spots (up to several times $10^3$K).
Falgarone et al. (1995) showed that the resulting chemical column
densities were consistent with observations. Joulain et al. (1998)
used a Burgers-vortex model for the dissipation to explain other
features, such as the similarity of CH and CH$^+$ line centroid
velocities, when the number of dissipation regions on the line of
sight is large.

Decamp \& Le Bourlot (2002) approximated turbulence by a
stochastic 1D velocity field correlated in space and time, and
solved the continuity equation for the density of each species.
For a small reaction network, the abundances developed different
patterns on different scales, showing how species segregation can
occur in a turbulent flow. Species that do not normally peak at
the same time for a nonturbulent model can coexist in a turbulent
model, which means that the age of a region cannot be inferred
from the abundances. The issue of chemical differentiation is
important because it is well-established and unexplained in a
number of starless cores such as TMC-1, L134N, and L1544, and
observed as 1--10 AU variations in CO, H$_2$CO, and OH relative to
H$_2$ toward extragalactic background sources (Marscher, Moore \&
Bania 1993; Moore \& Marscher 1995; for more recent work, see Pan,
Federman \& Welty 2001; Rollinde et al. 2003; Stanimirovic et al.
2003). Although some of these variations may be attributed to
freeze-out of molecules on dust grains (Bergin et al. 2002), most
are probably also affected by time-dependent chemistry (Langer et
al. 2000). One important timescale is cloud contraction
(Shematovich et al. 2003), but another is turbulent cycling
between different physical conditions over a range of timescales
(Decamp \& Le Bourlot 2002). The observed variations at scales so
small that gravity is unimportant seem to favor turbulence.  The
calculations by Decamp \& LeBourlet also develop a bistable
chemical pattern in which dual equilibria are possible. However,
it is not known how many of the effects result from the imposed
artificial velocity field, which is not able to react to the
changing density field through pressure.

The only truly hydrodynamical turbulence simulation that included
a small but relevant chemical reaction subset was by Pavlovski et
al. (2002), who generalized previous work by Mac Low and coworkers
to include a large number of coolants and a wide range of possible
temperatures and densities. The reaction network in Pavlovski et
al. included H, C, O, and five molecules formed from them, and it
included H$_2$ formation on grains. They followed the decay of
hypersonic turbulence without self-gravity in a dense ($10^6$
cm$^{-3}$), small ($10^{16}$ cm) region. Their surprising result
was that after H$_2$ dissociated in a shock, the molecules
reformed in a pattern of filaments, clumps, and diffuse gas yet
the abundances were fairly uniform after only 100 years.

There are probably other processes that produce localized regions
of hot chemistry. For example, ambipolar diffusion heating depends
on a high-order derivative of the magnetic field and so can
produce heating rates that exceed the average rate by orders of
magnitude on very small scales (Padoan, Zweibel, \& Nordlund
2000). Magnetic reconnection (Lazarian \& Vishniac 1999) is
another possibility for localized hot gas. The chemistry induced
by these processes has not yet been investigated.

\section{COSMIC RAY SCATTERING AND ACCELERATION IN A TURBULENT INTERSTELLAR MEDIUM}
\label{sect:cr}

Cosmic rays scatter off magnetic waves and MHD turbulence as the
particles propagate along magnetic field lines, and they generate
waves and turbulence if they stream much faster than the local
Alfv\'en speed.  Observations of cosmic rays at the Earth and of
radio and X-ray emissions produced by cosmic rays at their sources
and throughout the Galaxy suggest the ISM is filled with
irregularities on the scale of a particle gyroradius.  This ranges
from 1 AU to 1 pc for protons with energies between a few GeV and
the knee of the cosmic ray energy distribution at $10^{15}$ eV.
This section reviews the connection between cosmic rays and
turbulence. It is not known if the turbulence that scatters cosmic
rays has the same origin as the turbulence on larger scales, i.e.,
whether it is part of an energy cascade from these larger scales.
It could be generated locally, for example, by the cosmic rays
themselves or in other small scale instabilities (see also section
3, {\it Interstellar Turbulence I}), and it could be more similar
to scintillation turbulence, which has the same small scale
(Section \ref{sect:scint}) than molecular cloud turbulence, which
is presumably generated on much larger scales. Other reviews of
cosmic rays are in Cesarsky (1980), Berezinskii et al. (1990),
Drury et al. (2001), Jokipii (2001), Ptuskin (2001), and
Schlickeiser (2002).

The important point about cosmic rays from the perspective of this
review is that they scatter frequently, diffusing rather than
streaming through the Galaxy. Their total path length determined
from the ratio of secondary to primary nuclei is equivalent to
$10^4$ Galactic disk crossings at GeV energies (Engelmann et al.
1990, Higdon \& Lingenfelter 2003). Their flight time determined
from isotope ratios ranges from $\sim$15 My for a leaky-box model
with a small halo (Simpson \& Garcia-Munoz 1988, Yanasak et al.
2001) to $10^8$ years for a diffusion model in a $\sim$5 kpc halo
(Ptuskin \& Soutoul 1998). They scatter so frequently that their
flux at the Earth is isotropic to within $5\times10^{-4}$ (Cutler
\& Groom 1991, Munakata et al. 1997), even though there are only a
handful of nearby sources (i.e., Geminga, Vela, Lupus Loop, Loop
III---see Ptuskin \& Soutoul 1998, Jones et al. 2001, Taillet \&
Maurin 2003).

The scattering distance, $\lambda$, is the step size in a random
walk over the Galactic thickness $H$ (Achterberg, Blandford \&
Reynolds 1994). For $N$ random isotropic scatterings,
$H=N^{1/2}\lambda$, so for total distance travelled $vt=N\lambda$
with $v\sim c$ particle speed, we get $\lambda=H^2/vt=0.2 \;{\rm
pc} \left(H/{\rm kpc}\right)^2/\left(t/15 \;{\rm My}\right)$ at
GeV energies. This is a very small distance on galactic scales,
but still much larger than a gyroradius,
$R_G=\gamma\beta_r\sin\alpha mc^2/\left(ZeB_0\right)$ for rest
mass $m$, relativistic $\beta_r=v/c$, and
$\gamma=\left(1-\beta_r^2\right)^{-1/2}$, magnetic field strength
$B$(Gauss) and charge $Ze$ with $e=4.803\times10^{-10}$ esu.  In
the expression for $R_G$, the angle of particle motion with
respect to the field line is $\alpha$. Taking $\beta_r\sim1$ and
$\sin\alpha\sim1$, we get $R_G\sim 3.3\times10^{12}E_{\rm GeV}/
\left(ZB_{\mu G} \right)$ cm for hadron particle energy $E_{\rm
GeV}$ in GeV and field strength $B_{\mu G}$ in microGauss. Thus,
the mean scattering distance for GeV energies is $\sim$10$^5$
gyroradii and the gyroradius has the same scale as the Solar
system.

The relative amplitude of the magnetic irregularities $\delta
B/B_0$ that have the size $R_G$ may be determined from the random
walk of angular deflections, $\Delta \phi$, which build up in
proportion to the square root of the number $N_G$ of gyrations as
$\Delta \phi\sim N_G^{1/2}\delta B/B$ for small $\delta B/B$. The
scattering frequency is $\nu\equiv v/\lambda=\Delta\phi^2/t$, and
the number of gyrations is $N_G=\Omega t$ for gyrofrequency
$\Omega=v/R_G$. Thus, $\nu=\Omega \left(\delta B/B\right)^2$ and
$\delta B/B_0=\left(R_G/\lambda\right)^{1/2}\sim2\times10^{-3}$
when $\lambda\sim$0.2 pc.  Evidently, magnetic fluctuations of
less than one percent on 0.1 AU scales can isotropize cosmic rays
with GeV energies; in general, a wide spectrum of fluctuations is
needed to scatter the full energy range. This level of
irregularity is much smaller than observed on $>10$ pc scales from
pulsar dispersion measures, which suggest $\delta B\sim5\mu$G and
$<B>\sim1.5\mu$G (Rand \& Kulkarni 1989). It is also smaller than
the integrated fluctuations found by scintillation on scales less
than $3.5$ pc, which give $\delta B\sim0.9\mu$G and $<B>\sim3\mu$G
(Minter \& Spangler 1997), and smaller than $\delta B/B\sim1$ from
optical polarization (Heiles 1996, Fosalba et al. 2002). A
spectrum of magnetic energy fluctuations like the Kolmogorov or
Kraichnan spectra will make $\delta B/B$ much smaller on the scale
of the gyroradius than these observed field irregularities. Such a
spectrum will also give the required rigidity dependence for the
diffusion coefficient (Strong \& Moskalenko 1998). The rigidity of
a cosmic ray is $\gamma\beta_r mc^2/Ze$.

Cosmic rays can also scatter off strong magnetic irregularities
having a scale comparable to or larger than R$_G$ (Jokipii 1987).
Inverting the above expression and integrating over a spectrum of
waves $\propto\left(k/k_{min}\right)^{-q}$, the mean free path is
$\lambda\sim\left(B/\delta
B\right)^2R_G\left(k_{min}R_G\right)^{1-q}$ (Schlickeiser 1989).
For $\delta B/B\sim1$ and a particle gyroradius that fits inside
the compressed region, $R_G\leq k_{min}^{-1}$, this gives
$\lambda\sim R_G$. If the scattering arises in a network of strong
shocks in a supersonically turbulent medium, then the mean free
path for diffusion will be about the shock separation. Because the
gyroradius of GeV protons is less than the collision mean free
path for atoms and ions, collisional MHD models of compressible
turbulence do not apply to these small scales (see {\it
Interstellar Turbulence I} and Section \ref{sect:mirror} below).

\subsection{Cosmic Ray Scattering in Magnetic Waves}

The field line fluctuations that scatter cosmic rays could be
coherent waves, stochastic turbulence, or shock fronts. One common
view of waves is that they are weak fluctuations compared with the
mean field, with wave travel times much longer than an oscillation
period and a well-defined dispersion relation between frequency
and wavenumber. Particles in such a wave field interact
successively with many crests, travelling rapidly along the field
until their cumulative pitch angle deviation becomes large and
they begin to interact with a different wave train.  A turbulent
medium is not so regular but particle interactions with a broad
wave spectrum can be similar as long as the particle sees a
magnetic perturbation with the right wavelength after each
gyration. In both cases the field can be treated as static for the
fast-moving particles.

Much of the history of cosmic ray theory is based on observations
of the solar wind, which contains a mixture of weak and strong
fluctuations, depending on frequency (Dr\"oge 1994; Goldstein,
Roberts \& Matthaeus 1995). Perhaps $\sim$15\% of it is in the
form of slab-like Alfv\'en waves ($k_\bot\sim0$) and $\sim$85\% is
two-dimensional (2D) magnetic turbulence ($k_\|\sim0$, Bieber,
Wanner \& Matthaeus 1996). Here, $k_\bot$ and $k_\|$ are
wavenumbers perpendicular and parallel to the mean field. The
solar wind observations led to the quasilinear theory of cosmic
ray propagation (Jokipii 1966), which assumes the field is uniform
in space and unchanging in time over a particle gyration.
Interstellar applications of this theory may be appropriate at
high spatial frequencies where weak electron perturbations scatter
radio waves (Section \ref{sect:scint}), or near shock fronts where
weak waves on either side scatter relativistic particles back and
forth (Section \ref{sect:cracc}), or for waves that are excited by
the cosmic rays themselves (Chandran 2000a). The waves cause
cosmic rays to diffuse in both momentum and space (e.g.,
Schlickeiser 1994). Momentum diffusion means that velocities
change and isotropize; space diffusion means that cosmic rays
spread out slowly to more uniform densities. The two diffusion
coefficients are inversely related: Higher momentum diffusion
means lower spatial diffusion as the particles isotropize their
velocities more easily.

There are two types of waves that could be important for cosmic
ray scattering.  Alfv\'en waves are transverse oscillations with a
restoring force from field line tension and a frequency less than
the ion gyro frequency. Their dispersion relation between
frequency $\omega$ and wavenumber $k$ is $\omega^2=
v_A^2k^2\cos^2\theta$ for angle $\theta$ between the wave
propagation vector and the average field.  Their group and phase
velocity parallel to the mean field is the Alfv\'en speed
$v_A=B/\left(4\pi\rho_i\right)^{1/2}$ for ion density $\rho_i$.
When thermal pressure with sound speed $a$ contributes to the
restoring force there are also magnetosonic waves with the
dispersion relation
\begin{equation}\omega^2=0.5k^2\left(a^2+v_A^2\pm
\left[\left(a^2+v_A^2\right)^2-
\left(2av_A\cos\theta\right)^2\right]^{1/2}\right).\end{equation}
The positive and negative signs are for fast and slow waves. For
oblique propagation of magnetosonic waves, when $\cos\theta\sim0$,
the phase speed of the fast mode can be large along the field
lines: $v_A/\cos\theta$ in the low $\beta$ limit ($\beta$ is the
ratio of thermal to magnetic pressures, $=2\left(a/v_A\right)^2$)
and $a/\cos\theta$ in the high $\beta$ limit. This speed is
important for cosmic rays because they also move quickly along the
field lines and will resonate with the magnetic fluctuations in
these waves. Oblique magnetosonic waves also have a magnetic
pressure that varies slightly along the field as a result of the
successive convergences and divergences of the velocity and field,
which are in phase for fast waves and out of phase for slow waves.
These field line changes create magnetic mirrors that enhance the
scattering (Schlickeiser \& Miller 1998, Ragot 2000).

The condition for resonance between a cosmic ray and a wave is
(Jokipii 1966, Hall \& Sturrock 1967, Hasselmann \& Wibberenz
1968)
\begin{equation}
\omega-k_\|v_\|=n\Omega .\label{eq:resonance}
\end{equation}
This expression says that the Doppler shifted frequency of the
magnetic wave, as viewed by the particle moving parallel to the
field at speed $v_\|=v\mu\equiv v\cos\alpha$ for pitch angle
$\alpha$, equals an integer number, $n$, of the gyration
frequency. For an Alfv\'en wave, $\omega=k_\| v_A$, and for a fast
mode at low $\beta$, $\omega=k v_A$. The case $|n|\ge1$ is
gyroresonance.  For wave spectra that decrease with $k$ as a power
law, only the lowest-order resonances ($n=\pm1$) are important
(Cesarsky \& Kulsrud 1973).

Magnetosonic waves have an additional resonance for $n=0$ (Lee \&
V\"olk 1975, Fisk 1976, Achterberg 1981), when particles stay
between two crests by moving along the field at the parallel wave
speed.  Such a particle slows down when it hits the compressive
part of the wave ahead of it (by the mirror effect), and it speeds
up if the compressive part hits the particle from behind. As a
particle oscillates between crests, it can have a parallel
velocity that resonates with another wave moving faster along the
field, and then jump over to become trapped between two new
crests. For a broad wave spectrum, the particle jumps from one
wave to the next, gaining energy on average. This process is
called transit time acceleration of particles or transit time
damping (TTD) of waves (Miller 1997, Schlickeiser \& Miller 1998).
TTD increases the parallel component of cosmic ray momentum by
mirror scattering in the wave frame.  The mirror force depends on
the square of the perpendicular component of the velocity (Section
\ref{sect:mirror}), so TTD must operate with the pitch-angle
scattering of gyroresonance to maintain momentum isotropy (Miller
1997).

A particle that interacts with an Alfv\'en wave can resonate only
with the wavelength that is equal to the parallel gyrolength in
the Doppler shifted frame, $\left(v_A-v_\|\right)2\pi/\Omega$. A
particle that interacts with magnetosonic waves can respond to the
whole wave spectrum because it can choose the wave direction
$\theta$ where the parallel wave speed equals its own
(Schlickeiser \& Miller 1998, Schlickeiser \& Vainio 1999). In
practice, such a wave field may not have the appropriate $\theta$
because wave damping is larger at higher $\theta$.

TTD was originally thought to be such an important energy loss for
the waves that the fast mode would not be important for cosmic
rays (Schlickeiser 1994, Minter \& Spangler 1997, Tsap 2000).
Damping is large for high $\beta$ because the thermal particles
resonate with the waves and only highly parallel modes exist
(Holman et al. 1979, Foote \& Kulsrud 1979). However, $\beta<<1$
in the solar wind and in much of the cool dense phase of the ISM,
and observations of the solar wind show that oblique and fast
modes dominate over plane-parallel Alfv\'en waves (Tu, Marsch \&
Thieme 1989; Bieber, Wanner \& Matthaeus 1996; Matthaeus,
Goldstein \& Roberts 1990). Thus, TTD particle acceleration could
be important in parts of the ISM where $\beta$ is small. The waves
still have to be nearly isotropic to get effective scattering,
though, and they should not be damped by other processes, such as
ion-neutral or viscous friction.  Such damping is important in
most of the neutral medium where $\beta<1$ is otherwise favorable
for TTD (Kulsrud \& Pearce 1969, Felice \& Kulsrud 2001). This
leaves few places where wave damping is low enough that TTD might
be important.

Alfv\'en wave scattering is questionable too because of the
extreme anisotropy of the fluctuations that may cascade down from
pc scales to the gyroradius. Anisotropic turbulence produces tiny
perpendicular irregularities that average out over a particle
gyration (Chandran 2000a, 2001; Lerche \& Schlickeiser 2001; Yan
\& Lazarian 2002; Teufel, Lerche \& Schlickeiser 2003). The result
is a scattering coefficient that is many orders of magnitude less
than in the isotropic case. For this reason, Yan \& Lazarian
(2003a) reconsidered TTD scattering by fast modes at moderate
$\beta$, emphasizing the isotropy of the fast waves and
considering ISM regions where wave damping at particular cosmic
ray energies is relatively small. The resolution of this damping
problem for wave scattering is far from clear.

\subsection{Other Scattering Mechanisms}\label{sect:mirror}

The energy of fast cosmic ray gyrations is adiabatically invariant
during slow changes in the magnetic field amplitude, so an
increasing field puts more energy into the gyromotion while
removing it from the parallel motion. This change, combined with
momentum conservation, causes fast particles with high pitch
angles to bounce off converging field lines.  The mirror force is
$-M\nabla_\|B$ for magnetic moment $M=mv_\bot^2/2B$ with particle
mass $m$. This type of scattering involves very long wavelengths
parallel to the field (Ragot 1999, 2000; Felice \& Kulsrud 2001;
Lu et al. 2002). The mirror sources could also be molecular clouds
and clumps (Chandran 2000b).

Electric fields can also scatter cosmic rays. The flow of plasma
at speed $U$ transverse to a magnetic field generates an electric
field, ${\bf E}=-{\bf U}\times{\bf B}$, that can be important in
the parallel direction when the magnetic field has at least 2D
structure, and it is convected around by turbulent motions
(Fedorov et al. 1992, le Roux et al. 2002).

Transport transverse to the mean field occurs as individual field
lines wander (Jokipii \& Parker 1969, Bieber \& Matthaeus 1997,
Michalek \& Ostrowski 1998, Giacalone \& Jokipii 1999, K\'ota \&
Jokipii 2000). This is important for radial diffusion in the solar
wind, which has a spiral field (Chen \& Bieber 1993), for
diffusion in oblique shock fronts (Duffy et al. 1995), and for
diffusion in the vertical direction out of the Galaxy. Magnetic
fields with irregularities smaller than the gyroradius cause
particles to skip over field lines, giving anomalous diffusion
(Parker 1964; Chuvilgin \& Ptuskin 1993; Casse, Lemoine \&
Pelletier 2002; Otsuka \& Hada 2003; Erlykin et al. 2003). Braided
field lines can lead to sub-diffusion (Getmantsev 1963), when the
mean squared particle position in the cross field direction
increases as the square root of time, instead of directly with
time as in normal diffusion. Compound diffusion has both
cross-field diffusion and diffusion along the field lines (Kirk et
al. 1996). Cross diffusion requires a lot of perpendicular
structure to the field lines over a diffusion length for the
particle motion parallel to the field (Qin, Matthaeus \& Bieber
2002).  For ISM turbulence, the cross-field diffusion coefficient
is 0.1--0.2 of the parallel coefficient (Chandran \& Maron 2004a).

Shocks provide another site for cosmic ray scattering. Blandford
\& Ostriker (1980) considered an ISM made of cleared hot cavities
from supernovae and showed how the resulting shocks could scatter
and accelerate cosmic rays. Bykov \& Toptygin (1985) extended this
model to include secondary shocks that arise in the turbulence
caused by the supernovae. Klepach, Ptuskin \& Zirakashvili (2000)
included stellar wind shocks. The composite spectrum of cosmic
rays from an ensemble of shocks in a supersonically turbulent
medium was determined by Schneider (1993). In such a medium, the
summed energy distribution for cosmic rays from all of the shocks
is close to a power law, but there can be flat parts or bumps,
unlike the energy distribution from a single shock (see also
Achterberg 1990, Bykov \& Toptygin 1993).

\subsection{Cosmic Ray Acceleration}
\label{sect:cracc}

Cosmic rays gain energy as they scatter off randomly moving parts
of the ISM (Fermi 1949).  This is momentum diffusion and called
the second-order Fermi mechanism because the energy gained per
collision depends on the second power of the rms turbulent speed.
It is reminiscent of the thermalization of a star cluster with the
light particles gaining speed as they approach equilibrium with
the heavy particles. Scott \& Chevalier (1975) first applied this
mechanism to supernova remnants, using random motions inside the
remnant for scattering sites. Supernovae are a likely source for
cosmic rays because the cosmic ray energy density, $\sim$1 eV
cm$^{-3}$, divided by a $\sim$15 My lifetime in the Galaxy, is
about 5\% of the total supernova power (Baade \& Zwicky 1934,
Ginzburg \& Syrovatskii 1964).

Cosmic ray acceleration by momentum diffusion occurs throughout a
compressibly turbulent ISM, not just in supernova remnants
(Kulsrud \& Ferrari 1978, Ptuskin 1988, Dolginov \& Silant'ev
1990, Bykov \& Toptygin 1993, Webb et al. 2003, Chandran \& Maron
2004b). In a system of many shocks, particles that are trapped in
the weak magnetic turbulence of the ambient medium also get
accelerated every time these regions are compressed (Jokipii
1987).

Multiple shock crossings at the edge of a supernova remnant also
accelerate cosmic rays (see review in Blandford \& Eichler 1987).
If the field is parallel to the shock direction, then magnetic
turbulence created ahead of the shock by the outward streaming
particles (Wentzel 1974, Skilling 1975) scatters these particles
back into the shock where they encounter more turbulence. The
post-shock turbulence was formerly pre-shock turbulence that got
compressed and amplified (Schlickeiser, Campeanu \& Lerche 1993;
Vainio \& Schlickeiser 1998).  If the field is oblique to the
shock direction, then gyromotions (Jokipii 1987) and field line
wandering cycle particles through the front (Ragot 2001). Each
time a particle cycles through the front it gains energy from the
converging flow. This is called the first-order Fermi mechanism
(Fermi 1954) because the energy gained per crossing depends on the
first power of the shock velocity. About $10^{-3}$ of the incoming
thermal particles are injected into cosmic rays (Pryadko \&
Petrosian 1997; Kang, Jones \& Gieseler 2002; Bamba et al. 2003),
with an efficiency depending on obliqueness (Ellison, Baring \&
Jones 1995; Kobayakawa, Honda, Samura 2002). Generally, the
first-order mechanism dominates in strong shocks (Axford et al.
1977, Krymsky 1977, Bell 1978, Ostrowski 1994), although the
second mechanism is more important downstream than upstream
(Vainio \& Schlickeiser 1998).

Shock acceleration explains the cosmic ray energy spectrum up to
the ``knee'' at $\sim$10$^{15}$ eV.  The particles that stay in the
shock longest end up with the most energy.  The spectrum is a
distribution function in the number of shock crossings,
considering the continuous loss of particles that are trapped in
the downstream flow (e.g., Kato \& Takahara 2003).

Observations of the edge sharpness of supernova remnants suggest
the amplitude of MHD waves near the shock is $\sim$60 times the
average ISM value (Achterberg, Blandford \& Reynolds 1994).
Simonetti (1992) observed a factor of at least 10 in the magnetic
wave amplitude from Faraday rotation irregularities in a supernova
remnant compared with the adjacent line of sight. The X-ray
synchrotron emission from supernova remnants is direct evidence
for acceleration of relativistic electrons (Koyama et al. 1995,
Aschenbach \& Leahy 1999).

\subsection{Generation of Turbulence by Cosmic Rays}

A collection of particles with a high enough density cannot stream
along a magnetic field much faster than the Alfv\'en speed because
they generate magnetic irregularities that scatter them (Lerche
1967; Wentzel 1968a, 1969; Kulsrud \& Pearce 1969; Tademaru 1969).
The growth rate of the instability at wavenumber $k$ for isotropic
wave generation is (Cesarsky 1980)
\begin{equation}\Gamma\left(k\right)\sim\Omega
\left({{n_{CR}\left(k\right)}\over{n_i}}\right)
\left(-1+{{v_{stream}}\over{v_A}}\right).\end{equation} The first
term in the parenthesis represents the stabilizing effect of wave
damping from ion gyromotions.  The second term drives the
instability with a rate proportional to the ratio of the streaming
speed to the Alfv\'en speed.  The influence of collective effects
is in the ratio of the density of those cosmic rays that resonate
with wavenumber $k$ to the background density of thermal ions. The
size of the magnetic irregularity produced is about $2\pi$ times
the particle gyroradius ($\sim$10$^{13}$ cm for GeV energy and
$\mu$Gauss field), so more energetic particles make larger-scale
field distortions. These distortions are not at the bottom of a
turbulent cascade, so they should not be anisotropic (Chandran
2000a). They could cascade to give the ISM scintillation. The
upper limit to the cosmic ray energy that can generate waves
sufficient for their own scattering is around 100 to 1000 Gev
(Cesarsky 1980, Yan \& Lazarian 2003a). Much higher energies are
possible in a Galactic wind model where the boundary between
diffusion and advection depends on energy and the streaming
instability Landau damps in a highly ionized Galactic halo
(Ptuskin et al. 1997).

Hall (1980) suggested that waves generated by cosmic rays would
damp quickly and proposed instead that scintillation-scale
structures result from mirror and firehose instabilities in the
hot ISM phase.  Both require pressure anisotropies, which Hall
notes should arise at the level of 0.01-0.1 if the hot medium is
turbulent. The firehose instability requires
$P_\|-P_\bot>B^2/4\pi$ for parallel and perpendicular pressures
$P_\|$ and $P_\bot$ (Lerche 1966, Wentzel 1968b), and the mirror
instability requires
$P_\bot-P_|>\left(B^2/4\pi\right)\left(P_\|/B_\bot\right)$. These
conditions are satisfied for anisotropy
$|P_\|-P_\bot|/2P_\|\sim$0.01--0.1 if the plasma $\beta=P/P_{mag}$
is very large, as might be the case in the hot intercloud medium
if supernovae continuously sweep it free of gas and field.

\section{RADIO WAVE SCINTILLATION AS A DIAGNOSTIC FOR ISM TURBULENCE}
\label{sect:scint}

One of the earliest indications that the ISM is turbulent came
from scintillation observations of electron density fluctuations
on very small scales. These fluctuations cause diffraction and
refraction of radio signals from pulsars and a few extragalactic
sources. Diffraction broadens pulsar images, spreads out the pulse
arrival times, and narrows the frequency interval over which the
pulses have a coherent behavior. The relative motion of the
diffracting medium also modulates the pulsar intensity on a
timescale of minutes. Refraction from larger structures causes the
images to split or wander in position and to vary in intensity on
timescales of days to months. Scintillation effects like these
offer many diagnostics for electron density structures in the
ionized interstellar medium, including {\it H II}  regions, hot
bubble edges and the ionized and hot intercloud media. Still, the
origin of this turbulence is not clear. Because the scales are
very small, typically $10^{15}$ cm or below, it could be part of a
cascade from larger structures past the viscous length and down
into the collisionless regime of MHD turbulence (section 4.11 in
{\it Interstellar Turbulence I}), or it could be generated locally
by cosmic ray streaming or other small-scale instabilities or by
low-mass stellar winds and wakes (sections 3 and \ref{sect:cr} in
{\it Interstellar Turbulence I}). Here we review ISM
scintillations. More extensive reviews may be found in Rickett
(1977, 1990), Hewish (1992), and Cordes, Rickett \& Backer (1988).

\subsection{Theory of Diffraction and Refraction in the ISM}

The index of refraction for radio waves is (Nicholson 1983)
\begin{equation}\eta=1-4\pi n_er_e/k^2\label{eq:index}\end{equation}
with electron density $n_e$, classical electron radius
$r_e=e^2/\left(m_ec^2\right)=2.8\times10^{-13}$ cm, and wavenumber
of the radio signal $k=2\pi/\lambda$ for wavelength $\lambda$. The
phase change of a signal that passes through a clump of size
$\delta x$ with an excess index of refraction $\delta \eta$
compared with a neighboring average region is $k\delta x\delta
\eta$. The transmitted signal adds constructively to the
neighboring signal at a relative propagation angle $\delta \theta$
if the difference in their path lengths, $\delta x\sin\delta
\theta$, multiplied by $k$, equals the phase change. This gives
$\delta \theta\sim\delta \eta= 4\pi\delta n_er_e/k^2$ for small
$\delta \theta$. This phase change is random for each clump the
signal meets, so after $N=D/\delta x$ such clumps on a path length
$D$, the root mean square angular change of the signal from
diffraction is $\delta\theta_d=\delta \eta N^{1/2}$.

The most important clump size has a cumulative scattering angle
$\delta \eta\left(D/\delta x\right)^{1/2}$ equal to the clump
diffraction angle, $\delta\theta_d=\left(k\delta x\right)^{-1}$.
Evaluation of $\delta\eta$ requires some knowledge of how $\delta
n_e$ depends on $\delta x$. For a turbulent medium, this relation
comes from the power spectrum of electron density fluctuations,
\begin{equation}
P(\kappa)=C_n^2 \kappa^{-\alpha},\end{equation} where
$\kappa=1/\delta x$ and $\alpha=11/3$ for 3D turbulence with a
Kolmogorov spectrum (we use $\kappa$ here for the wavenumber of
electron density fluctuations to distinguish it from the
wavenumber $k$ of the radio radiation used for the observation).
The mean squared electron density fluctuation is obtained from the
integral over phase space volume, $P(\kappa)4\pi \kappa^2d\kappa$.
For logarithmic intervals, this is approximately $\delta
n_e^2=P(\kappa)4\pi\kappa^3 \sim 4\pi C_n^2 \delta x^{\alpha-3}$.
Thus, we have $\delta x\sim \left(16\pi\lambda^2 r_e^2
DC_n^2\right)^{-1/\left(\alpha-2\right)}$. Using the wave
equation, Cordes, Pidwerbetsky \& Lovelace (1986) got essentially
the same result with $4\pi^2$ replacing $16\pi$.  The full theory
more properly accounts for differences in the cumulative effects
of the line-of-sight and transverse density variations. A recent
modification considers power-law rather than Gaussian statistics
for the clump size distribution (Boldyrev \& Gwinn 2003).

For a reference set of parameters, e.g., GHz frequencies, kpc
distances, and typical $C_n^2\sim10^{-4}$ meters$^{-20/3}$, the
characteristic scale is $\delta x\sim10^{10}$ cm. This small size
implies that only pulsars and a few extragalactic radio sources
can produce detectible diffraction effects at radio frequencies.
The smallness compared with the collisional mean free path of
electrons also implies that the fluctuations have to be
collisionless (see {\it Interstellar Turbulence I}).

Many observable properties of small radio sources give
information, either directly or indirectly, about $\delta x$,
which can be used to infer the strength and power index of the
electron density fluctuation spectrum and its distribution on the
line of sight. For example, the logarithm of the visibility of an
interferometer is $-4\pi^2\lambda^2r_e^2s^{\alpha-2}SM$ where $s$
is the baseline length and $SM=\int_0^D C_n^2dz$ is the scattering
measure. Spangler \& Cordes (1998) found $\alpha\sim3.65\pm0.08$
using this baseline dependence for 4 pulsars.  Very long baseline
interferometry (VLBI) observations of pulsar image sizes measure
angular broadening as a function of frequency (Lee \& Jokipii
1975a; Cordes, Pidwerbetsky \& Lovelace 1986) and determine $SM$
(Spangler et al. 1986). For the reference parameters, $\delta
\theta_d\sim0.15$ mas.

Angular broadening gives a spread in path lengths for the radio
waves, and this corresponds to a spread in arrival times of
pulses, $\tau=D\delta\theta_d^2/\left(2c\right)$ (Lee \& Jokipii
1975b). Such pulse broadening scales as
$\lambda^{2\alpha/\left(\alpha-2\right)} {\rm
SM}^{2/\left(\alpha-2\right)}D$, so the frequency dependence gives
$\alpha$ and the absolute value gives $SM$ for an assumed screen
depth $D$. The flux density is correlated only over a small range
of frequencies, $\delta \nu_d\sim\left(2\pi \tau\right)^{-1}$
(Salpeter 1969, Lee \& Jokipii 1975b).

Pulsar amplitudes vary on a timescale $\delta t_d=\delta
x/v_\bot\sim$ several minutes as a result of pulsar transverse
motions $v_\bot\sim100$ km s$^{-1}$ (Cordes 1986). Using the
relations above, this can be rewritten $\delta
t_d\sim\left(cD\delta\nu_d \right)^{1/2}/\left(2\pi^{1/2}\nu
v_\bot\right)$. If pulsar distances and proper motions are also
known, along with the dispersion measure, then the distribution of
scattering material on the line of sight can be modeled (Harrison
\& Lyne 1993, Cordes \& Rickett 1998).

Longer time variations (Sieber 1982) result from the changing
refraction of radio signals in moving structures that have the
diffraction angular size $\delta \theta_d$ or larger (Rickett,
Coles \& Bourgois 1984; Blandford \& Narayan 1985). This angular
size corresponds to a physical size for the refracting elements
$\delta x_r=D\delta\theta_d$. For typical $\delta\theta_d\sim0.1$
mas, $\delta x_r\sim1.5\times10^{12}$ cm at $D=1$ kpc.  The
corresponding refraction scintillation time is $\delta t_r=\delta
x_r/v_\bot\sim$ days. Time variations over months are also
observed from larger interstellar structures.

The relative rms amplitude of the source is the modulation index,
$m=<\left(I-<I>\right)^2>^{1/2}/<I>$. For diffraction this can be
100\%, but for refraction it is typically less than $\sim$30\%
(Stinebring \& Condon 1990, LaBrecque et al. 1994, Stinebring et
al. 2000).  The modulation index depends on the strength of
scattering, which involves a length scale equal to the geometric
mean of the scale for angular broadening and the dominant scale
for electron density fluctuations, $\left(D\delta\theta_d\delta
x\right)^{1/2}$.  This is the Fresnel length,
$r_F=\left(D/k\right)^{1/2}$, which is the transverse size of some
object at distance $D$ that is just small enough to show
diffraction effects.  When $r_F/\delta x_d$ is high, there are
many diffracting elements of size $\delta x_d$ inside each
refracting element of size $D\delta\theta_d$, so diffraction is
strong. For weak diffraction, the total phase change on the line
of sight is small and the dominant clump size is the Fresnel
length itself (Lovelace et al. 1970, Lee \& Jokipii 1975c,
Rickett 1990). In this case, the modulation index scales
approximately with $\left(r_F/\delta x_d\right)^{\alpha-2}<1$
(Lovelace et al. 1970). The modulation index measures interstellar
properties in this weak limit because then $m^2\propto
\lambda^{\left(\alpha+2\right)/2} D^{\alpha/2}C_n^2$, giving a
diagnostic for $C_n^2$ and $\alpha$ (Rickett 1977). Small $m$
usually corresponds to high frequencies, where diffraction is
relatively unimportant.

Time variations can be visualized with a dynamic spectrum, which
is a gray scale plot of intensity on a coordinate system of time
versus frequency (Figure \ref{fig:dynamspect}). Diffraction alone
gives a random dynamic spectrum from the motion of unresolved
objects of size $\delta x$, but larger refractive structures,
which disperse their radio frequencies over larger angles,
$\delta\theta\sim-2\theta\delta\nu/\nu$ (from the above expression
$\delta \theta\sim 4\pi\delta n_er_e/k^2$), sweep a spectrum of
frequencies past the observer (Hewish 1980). The intensity peaks
on a dynamic spectrum (averaged over many pulses) then appear as
streaks with slope $dt/d\nu=-2D\delta\theta_d/\left(\nu
v_\bot\right)$ (Cordes, Pidwerbetsky \& Lovelace 1986); this can
be used as a diagnostic for $v_\bot$. The right-hand part of
Figure \ref{fig:dynamspect} is the 2D power spectrum of the
dynamic spectrum. Stinebring et al. (2001) and Hill et al. (2003)
suggest that the arcs arise from interference between a central
image and a faint scattering halo 20--30 times larger.

Strong refraction can result in multiple images that produce
interference fringes on a dynamic spectrum (Cordes \& Wolszczan
1986; Rickett, Lyne \& Gupta 1997). Multiple images imply that
refraction, which determines the image separation, bends light
more than diffraction, which determines the size.  Because
large-scale fluctuations dominate refraction, this observation
implies either $\alpha>4$, so that small wavenumbers dominate
(Cordes, Pidwerbetsky \& Lovelace 1986; Romani, Narayan \&
Blandford 1986), or there is additional structure on $10^{12}$ cm
scales that is not part of a power law power spectrum (Rickett,
Lyne \& Gupta 1997).

\subsection{Observations}

Radio scintillation has been used to determine many properties of
electron density fluctuations in the ISM. Diffraction and
refraction come from the same electrons, so the ratio of their
strengths is proportional to the ratio of the amplitudes of the
electron density fluctuations on two different scales. From this
ratio Armstrong, Rickett \& Spangler (1995) derived the power
spectrum of the fluctuations spanning $>6$ orders of magnitude in
scale. They determined $C_n^2=10^{-3}$ m$^{-20/3}$ and
$\alpha=11/3$ between $10^6$ cm and $10^{13}$ cm. They also
suggested from rotation measures (RM$=\int_0^Dn_eB_\|dz$) that the
same power law extends up to $10^{17}$ or $10^{18}$ cm.

Cordes, Weisberg \& Boriakoff (1985) found $\alpha=3.63\pm0.2$
using the relation between $\delta\nu_d$ and $\lambda$. They
mapped the spatial distribution of $C_n^2$ from $\delta\nu_d$
observations of 31 pulsars, suggesting the Galaxy contains both
thin and thick disk components. For the thin component,
$C_n^2\sim10^{-3}-1$ m$^{-20/3}$ assuming $\alpha=11/3$ and for
the thick component, with $H>0.5$ kpc, $C_n^2\sim10^{-3.5}$
m$^{-20/3}$.  The rms level of electron density fluctuations at
the dominant scale $\delta x\sim10^{10}$ cm, which comes from the
integral over the power spectrum (see above), was found to be
$<\delta n_e^2>^{1/2}\sim5\times10^{-6}$ cm$^{-3}$ for the high
latitude medium, for which $<n_e>\sim0.03$ cm$^{-3}$ from
dispersion measures, and $<\delta
n_e^2>^{1/2}\sim10^{-4.2}-10^{-3.3}$ cm$^{-3}$ for the low
latitudes.

Bhat, Gupta \& Rao (1998) observed 20 pulsars for three years to
average over refraction variations and determined $C_n^2$ in the
local ISM from $\delta\nu_d$. They observed an excess of
scattering material at the edge of the local bubble, which also
produced multiple images of one of these pulsars (Gupta, Bhat \&
Rao 1999). The edge of the local bubble may also have been seen by
Rickett, Kedziora-Chudczer \& Jauncey (2002) as a source of
scattering in a quasar. Bhat \& Gupta (2002) found a similar
enhancement at the edge of Loop I, where $SM\sim0.3$ pc
m$^{-20/3}$ and the density enhancement is a factor of $\sim$100
over the surrounding gas. They also found excess scattering for
more distant pulsars from the Sagittarius spiral arm.

Lazio \& Cordes (1998a,b) used the angular sizes of radio sources
and other information on lines of sight to the Galactic center and
outer Galaxy to suggest that the scattering material is associated
with the surfaces of molecular clouds. Ionized cloud edges were
also suggested by Rickett, Lyne \& Gupta (1997) to explain
multiple images. Spangler \& Cordes (1998) observed $\delta
\theta_d$ from small sources behind six regions in the Cygnus OB1
association and found an excess in SM that was correlated with the
emission measure, indicating again that scattering is associated
with {\it H II} regions. This result is consistent with the high
cooling rate required by the dissipation of this turbulence, which
implies high temperatures (Zweibel, Ferriere \& Shull 1988;
Spangler 1991; Minter \& Spangler 1997).

The relation $v_\bot\sim\left(cD\delta\nu_d \right)^{1/2}/
\left(2\pi^{1/2}\nu\delta t_d\right)$ combined with pulsar
distances and proper motions led Gupta (1995) to determine a 1 kpc
scale height for the scattering layer from long-term observations
of 59 pulsars. Cordes \& Rickett (1998) also used this method to
find that scattering is rather uniformly distributed toward two
pulsars, but for three lines of sight it was concentrated toward
the pulsar, including Vela where the supernova remnant is known to
contribute to the scattering (Desai et al. 1992); six other
pulsars in that study had significant scattering from either a
foreground spiral arm or {\it H II} regions.

Bhat, Gupta \& Rao (1999) compared $\delta\theta_d$ obtained from
$\delta\nu_d$ with the angular size of the refraction pattern,
$\delta\theta_r\sim\left(v_\bot/D\right)\left(dt/d\nu\right)$ from
dynamic spectra. They showed that
$\delta\theta_r/\delta\theta_d<1$ for all 25 pulsars that had this
data, implying that diffraction dominates refraction and therefore
$\alpha<4$.  They also compared $C_n^2$ from $\delta\theta_r$ on
the refractive scale with $C_n^2$ from $\delta\nu_d$ on the
diffractive scale to determine the slope of the power spectrum
directly (as did Armstrong et al. 1995).  They found the
Kolmogorov value $\alpha=11/3$ to within the accuracy in most
cases, but six pulsars, mostly nearby, gave slightly higher
$\alpha\sim3.8$.  The longest-term variations had significantly
more power than an extrapolation of the Kolmogorov spectrum,
however, and the modulation indices were large, leading them to
suggest an additional component of scattering electrons at scales
of $10^{14}$ to $10^{15}$ cm.  The same data led Bhat, Rao \&
Gupta (1999) to derive $<C_n^2>\sim10^{-3.8}$ m$^{-20/3}$ and
$<r_F/\delta x_d>\sim45$, indicating strong scattering.

Lambert \& Rickett (2000) looked at the correlation between the
modulation index for long-term variations from refraction and the
relative decorrelation bandwidth, $\delta \nu_d/\nu$, which comes
from diffraction. The correlation at 610 MHz for 28 sources fit
Kolmogorov scaling ($\alpha=11/3$) better than a shock-dominated
model ($\alpha=4$) model but at 100 MHz the Kolmogorov fit was not
as good.  They suggested that the minimum turbulent length was
large, $10^{10}$ to $10^{12}$ cm instead of $<10^9$ cm (Armstrong,
Rickett \& Spangler 1995). An excess of electron density structure
at large scales could also explain the discrepancy.

Stinebring et al. (2000) monitored the modulation index of 21
pulsars for five years and found low values ($<50$\%) that bracket
$3.5<\alpha<3.7$ with no perceptible inner (smallest) scale for
most of the pulsars. For the Crab, Vela, and four others, enhanced
modulation indices were consistent with an inner scale of
$10^{10}$ cm. Rickett, Lyne \& Gupta (1997) suggested this excess
scattering could be from AU-sized ionized clumps at cloud edges.

An inner scale of $5\times10^6-2\times10^7$ cm was found by
Spangler \& Gwinn (1990) after noting that interferometer
baselines shorter than this had scintillation with $\alpha\sim$4
whereas longer baselines had $\alpha$ near the Kolmogorov value,
$-11/3$. They suggested the inner scale is caused by a lack of
turbulence smaller than the ion gyroradius, $v_{th}/\Omega$ for
thermal speed $v_{th}$, or by the ion inertial length,
$v_A/\Omega$, whichever is larger.  This value of the inner scale
is consistent with an origin of the scattering in the warm ionized
medium or in {\it H II} regions, but not in the hot coronal
medium, which has much larger minimum lengths.

Shishov et al. (2003) studied pulsar PSR B0329+54 over a wide
range of frequencies and found a power spectrum for electron
density fluctuations with a slope of $-3.5\pm0.05$ for lengths
between $10^8$ cm and $10^{11}$ cm. This spectrum is expected for
turbulence in the Iroshnikov-Kraichnan model (see {\it
Interstellar Turbulence I}). Shishov et al. noted how other lines
of sight though the galaxy gave different spectra and suggested
that the nature of the turbulence varies from place to place. They
also found refraction effects on a scale of $3\times10^{15}$ cm
corresponding to electron density fluctuations of strength $\delta
n_e\sim10^{-2}$ cm$^{-3}$ and suggested these were neutral clouds
with an overall filling factor of $\sim$0.1.

Scattering image anisotropy suggests anisotropic turbulence (see
{\it Interstellar Turbulence I}). Lo et al. (1993) observed 2:1
anisotropy in images of Sgr A$^*$, Wilkinson, Narayan \& Spencer
(1994) found scale-dependent anisotropic images of Cygnus X-3,
Frail et al. (1994) observed 3:1 anisotropy in scattering of light
from Galactic center OH/IR stars, whereas Trotter, Moran \&
Rodriguez (1998) observed axial ratios of 1.2--1.5 for quasar
light that scatters through a local {\it H II} region. A ratio of
4:1 was found for another quasar by Rickett, Kedziora-Chudczer \&
Jauncey (2002). Spangler \& Cordes (1998) observed anisotropic
scattering with axial ratio of 1.8 surrounding the Cygnus OB
association and Desai \& Fey (2001) observed axial ratios of
$\sim$1.3 toward the same region. The actual anisotropy of local
fluctuations cannot be determined from these observations because
many different orientations blend on the line of sight (Chandran
\& Backer 2002)

Extreme scattering events are observed in some extragalactic radio
sources and a few pulsars. Their modulation is strong, $\ge50$\%,
they can last for several months, and they have light profiles
that are flat-bottom with spikes at the end, or smooth bottom with
no spikes (Fiedler et al. 1994). They may result from supernova
shocks viewed edge-on (Romani, Blandford \& Cordes 1987) or
ionized cloud edges in the Galaxy halo (Walker \& Wardle 1998).
Some appear correlated with the edges of local radio loops
(Fiedler et al. 1994, Lazio et al. 2000), in support of the shock
model.  The actual scattering process could be a combination of
refractive defocusing, during which an intervening electron cloud
produces a lens that diverges the light and makes the source
dimmer (Romani et al. 1987, Clegg et al. 1998), or stochastic
broadening by an excess of turbulence in a small cloud (Fiedler et
al. 1987). Lazio et al. (2000) found an excess $SM=10^{-2.5}$ kpc
m$^{-20/3}$ associated with an event, corresponding to
$C_n^2\sim10^7\xi^{-1}\left(D/100\;{\rm pc}\right)^{-1}$
m$^{-20/3}$ for ratio $\xi$ of the line-of-sight extent ($D$) to
the transverse extent.  Fiedler et al. (1994) suggested $<\delta
n_e^2>^{1/2}\sim10^2/\xi$ cm$^{-3}$.  This high level of
scattering along with an observed increase in angular size during
the brightness minimum (Lazio et al. 2000) suggests the scattering
cloud is not part of a power spectrum of turbulence but is an
additional AU-size feature (Lazio et al. 2000).

\subsection{Summary of Scintillation}

The amplitude, slope, and anisotropy of the power spectrum of
interstellar electron density fluctuations have been observed by
scintillation experiments. Estimates for the inner scale of these
fluctuations range from $10^7$ cm to $10^{10}$ cm or more. The
shorter of these lengths is about the ion gyroradius in the warm
ionized medium (Spangler \& Gwinn 1990). This is much smaller than
the mean free path for electron collisions so the fluctuations are
collisionless (see {\it Interstellar Turbulence I}). A lower limit
to the largest scale of $\sim$10$^{18}$ cm was inferred from
rotation measures assuming a continuous power law for
fluctuations, but this assumption is uncertain. The slope of the
power spectrum is usually close to the Kolmogorov value, $-11/3$,
and distinct from the slope for a field of discontinuities, which
is $-4$. Deviations from the Kolmogorov slope appear in some
studies, and may be from a large inner scale, $>>10^9$ cm, an
excess of scattering sites having size $\sim$1--10 AU, or a
transition in the scaling properties of the turbulence.

Scintillation arises from both the low-density, diffuse, ionized
ISM and the higher-density {\it H II} regions, ionized cloud
edges, and hot shells, where the amplitude of the power spectrum
increases. On average, the relative fluctuations are extremely
small, $<\delta n_e^2>^{1/2}/<n_e>\sim10^{-2}$ on scales that
dominate the diffraction at GHz radio frequencies, $\delta
x\sim10^{10}$ cm (Cordes et al. 1985). For a Kolmogorov spectrum,
they would be larger on larger scales in proportion to $L^{2/3}$;
if the spectrum is continuous up to pc scales, the absolute
fluctuation amplitude there would be $\sim$1 (Lee \& Jokipii
1976). Anisotropy of scattered images is at the level of 50\%,
which is consistent with extremely large intrinsic anisotropies
($\times1000$) from MHD turbulence (sections 4.11 and 4.13 in {\it
Interstellar Turbulence I}) if many orientations on the line of
sight are blended together.

\section{Summary and Reflections}

The turbulence that is observed directly on resolvable scales in
the ISM also has important effects on very small scales, down to
the level of atomic diffusion, mixing, and viscosity, and
continuing far below to the thermal ion gyroradius.  The
resolvable turbulence in the neutral medium was reviewed in the
first part of this series ({\it Interstellar Turbulence I}), along
with general theory and simulations. The smaller scale
implications were reviewed here.

Turbulence helps to disperse the elements made in supernovae and
other stellar sources by stretching and folding the contaminated
gas until the gradient length approaches the collision mean free
path. Then atomic diffusion does the final step of interatomic
mixing. This mixing and homogenization may occur partly at the
source, in the shock fronts and contact discontinuities around the
expanding flow following dynamical instabilities, and it may occur
partly in the ambient ISM after the expansion subsides.

Observations suggest that the dispersion in elemental abundances
among stars inside clusters and field stars of the same age, and
the abundance differences between HII regions and diffuse clouds,
are only a few percent up to perhaps 30\%. Observations of
clusters also suggest that most of the mixing occurs within
several times $10^8$ years. This homogenization has to occur
between the injection scale of supernovae and superbubble
explosions and the star formation scale containing on the order of
a solar mass. The corresponding range of spatial scales is
$\sim100$ for gas at the ambient density. Homogenization cannot be
significant on scales much larger than a superbubble, because then
it would diminish the galactic radial abundance gradient over a
Hubble time.  These gradients enter the problem in another way too
because turbulence mixes the gas over galactic radius to increase
the elemental dispersion locally at the same time as it mixes this
gas locally and leads to homogenization.

While the distribution of elemental abundances appears to be
fairly narrow, with a dispersion on the order of 10\%, it may also
have a fat tail caused by occasional odd stars and diffuse clouds
with very different abundances. This tail is reminiscent of other
fat tails in the distribution functions for turbulent media, and
these other tails seem to originate with intermittency. In the
case of elemental dispersion, this means that the homogenization
process is spotty so some regions survive for long periods of time
with very little mixing.

Several methods for studying turbulent mixing have been employed.
Because of the wide range of scales involved, these studies often
employ simplifying assumptions that appear to capture the
essential physics. They include artificial stochastic velocity
fields and closure methods using moment equations. Direct
numerical simulations of ISM elemental dispersion have only just
begun.  The theory suggests that a turbulent medium mixes passive
scalars like elemental abundances faster than a Gaussian velocity
field because of the long-range correlations that are associated
with turbulence.

Turbulence affects interstellar chemistry by mixing regions with
different properties, heating the gas intermittently on the
viscous scale, and enhancing ion-neutral collisions in regions
with strong magnetic field gradients. Turbulent mixing is more
far-reaching than diffusive mixing because of long-range
correlated motions in a turbulent flow. Such mixing spreads out
each chemical species over a large radial range inside a cloud.
Turbulent heating promotes temperature-sensitive reactions inside
otherwise cold clouds. Applications to the formation of CH$^+$ and
OH in diffuse clouds look promising.  Chemical reaction networks
that include these turbulent processes are only beginning to
understand some of the implications, and direct simulations of
chemistry in a turbulent medium are limited to only a few studies
so far.

The scattering and acceleration of cosmic rays depends strongly on
the presence of turbulence in the ISM. The scale for this
turbulence is the gyroradius, which is less than the collisional
mean free path for most cosmic rays, which have energies less than
1 GeV.  Thus, scattering relies entirely on magnetic
irregularities in collisionless plasma turbulence. Cosmic ray
acceleration is by two processes, both of which involve
turbulence: The first-order Fermi mechanism accelerates cosmic
rays by cycling them through shock fronts where they pick up a
relative velocity kick comparable to the shock speed at each
passage.  This cycling occurs because magnetic irregularities on
each side of the front scatter the out-streaming cosmic rays back
into the front.   The second-order Fermi mechanism accelerates
cosmic rays through turbulent diffusion. Each collision with a
randomly moving magnetic irregularity turns the cosmic ray around
with a reflection speed in the moving frame that is comparable to
the incident speed.  Over time, the result is a transfer of energy
from the turbulence to the cosmic rays.

Cosmic ray scattering occurs in several ways. Fast moving
particles can interact weakly but resonantly with numerous
magnetic irregularities, and successive interactions of this type
can randomly change the pitch angles of their helical motions
along the field. This eventually leads to a complete reversal in
the direction of motion. Cosmic rays can also interact strongly
with large magnetic irregularities, as would occur in shock fronts
and at the edges of dense cloud complexes. These interactions
change the particle directions significantly each time.  The
nature of cosmic ray diffusion in both momentum and space is not
understood well because the structure and strength of the
important magnetic irregularities are not observed directly.  If
MHD turbulence is highly anisotropic on the scale of cosmic ray
scattering, with transverse irregularities much stronger than
parallel as suggested by theory, then resonant scattering
processes during motions along the mean field can be very weak.
The structure of magnetic waves below the mean free path is also
unclear, as the usual fast and slow magnetosonic modes do not
exist.

Cosmic rays can also generate turbulence by streaming
instabilities following particle-wave resonances or by fire-hose
and mirror instabilities that operate even without resonances. The
expected anisotropy of Alfv\'en wave turbulence diminishes the
first of these mechanisms significantly, however.  The others are
problematic because they require strong cosmic ray pressure
anisotropies.  As a result, the impact of cosmic rays on
turbulence is currently not understood.

Radio wave scintillation is indirect evidence for interstellar
plasma turbulence. These radio observations span a very wide range
of spatial scales through a combination of diffraction and
refraction effects.  The scales are mostly below the collision
mean free path, and they are far below the limits of angular
resolution. The first of these limits makes it difficult to
understand the origin of the density irregularities. Unlike the
turbulence that scatters cosmic rays, which requires only magnetic
irregularities and no density structure, the turbulence that
causes scintillation requires small-scale density irregularities
in the ionized medium.  The associated magnetic irregularities are
not observed, and the connection to cosmic ray scattering is
unclear, even though the length scales are about the same. The
origin of density structures below the collision mean free path is
unknown. Atomic diffusion should smooth them out on a sound
crossing time unless magnetic field irregularities hold them in
place. In that case they could be the result of slight temperature
variations, with the cooler regions having higher electron
densities, all divided up and mixed together by transverse
magnetic motions. The nature of these motions below the mean free
path is unclear, because, as mentioned above, the usual fast and
slow compressional modes do not exist, nor do the usual thermal
and pressure-regulated processes.

The second of the two limits on spatial scale imply that the
geometrical properties of scintillation turbulence are difficult
to observe. The scintillation is clearly anisotropic, but whether
it is in sheets or filaments, for example, is unknown.

The importance of scintillation observations for studies of ISM
turbulence is that they give the power spectrum of electron
density fluctuations fairly accurately.  This is usually close to
the Kolmogorov spectrum of incompressible turbulence. Rarely are
the spectra so steep that the medium can be interpreted as a
superposition of sharp edges, like shock fronts. One recent
observation with fairly high precision obtained the relatively
shallow Iroshnikov-Kraichnan power spectrum, leading to the
conjecture that the slope varies from region to region. As
discussed in {\it Interstellar Turbulence I}, this variation may
arise from a variation in the relative strength of the magnetic
field compared to the turbulent motions, with the
Iroshnikov-Kraichnan spectrum present in regions of relatively
strong fields.

Observations suggest that scintillation arises in a distributed
fashion from the ambient ionized medium and also from discrete
high-density places like the edges of local bubbles, ionized
molecular clouds, and HII regions.  These discrete regions are
likely to be highly turbulent and they also have a juxtaposition
of hot and cool gas, which is necessary for isentropic mixing and
electron density structure.

There are evidently many uncertainties in the nature of ISM
turbulence on small scales, even though the evidence for this
turbulence is pervasive. Part of the problem is that none of the
features of this turbulence have been observed directly: not the
densities, magnetic fields, temperatures, or motions. Still, the
density irregularities are revealed indirectly through
scintillation, the magnetic field irregularities through cosmic
ray scattering, the temperature fluctuations through chemistry,
and the motions through elemental and chemical mixing. An
additional problem is that many small scale effects of ISM
turbulence rely on details of the theory that are independent of
the usual scaling relations, such as viscous heating and elemental
diffusion, which arise at the bottom of the cascade in the neutral
medium.  Turbulence in the ionized medium is also below the
collisional mean free path, where pressure and thermal effects are
relatively unimportant. Moreover, the small scale ISM processes
are often strongly dependent on the large scale processes, such as
turbulent shock formation, energy and metal injection, and
galactic-scale gradients.  This means that direct simulations of
small scale turbulence are impossible without simplifying
assumptions about the large-scale medium -- assumptions that
require more knowledge about ISM turbulence on the large scale
than is presently available.

While the observations and theory of ISM turbulence have come a
long way from the first efforts in the 1950s, the details of this
new information have led to a growing awareness that the complete
problem is far too large to solve any time soon.  We rely on
future generations of astronomers and physicists to continue this
work, and hope that they find this field as intriguing and
challenging as we do today.

\section{\sc Acknowledgments} 
\begin{figure}
\plotone{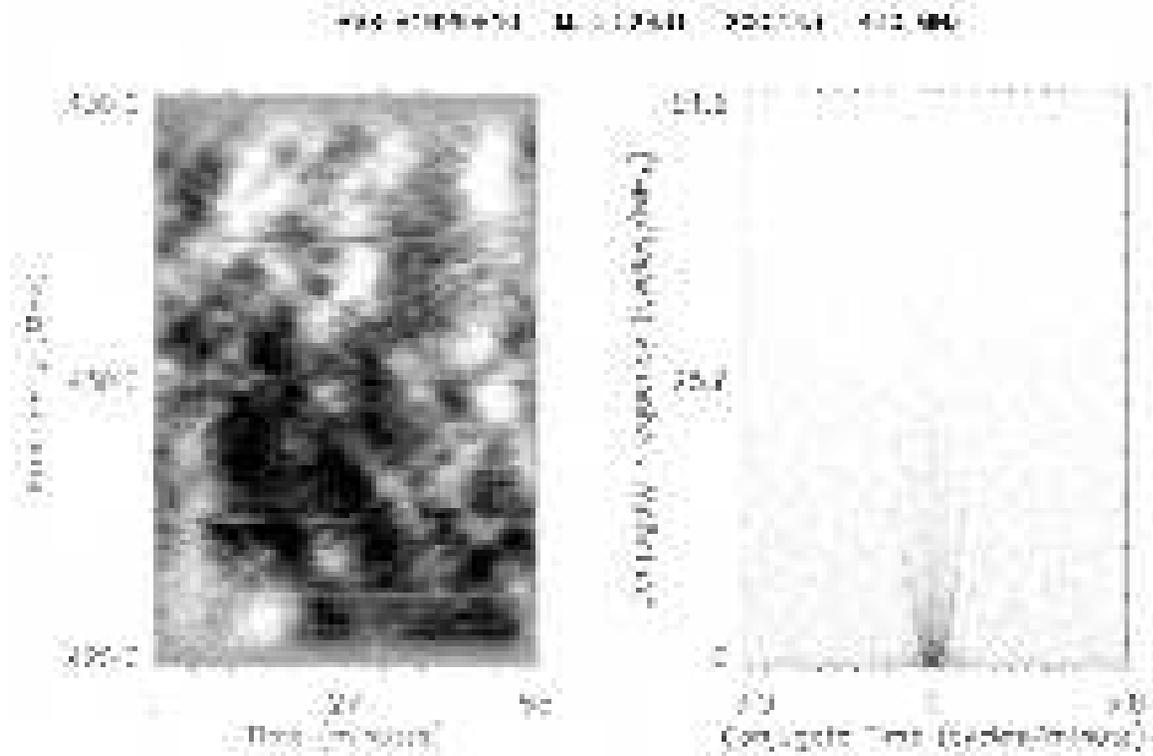} 
\caption{Dynamic spectrum of
pulsar PSR B1929+10 ({\it left}) plotting flux density linearly
with grayscale. The vertical columns are spectra, and many spectra
are aligned horizontally in time. Intervening fluctuations in the
electron density cause the signal to drift in both frequency and
arrival time. A 2D Fourier transform of the dynamic spectrum is
shown on the right. The crisscross pattern in the dynamic spectrum
causes the parabolic boundary in the Fourier transform
distribution. The grayscale for the secondary spectrum is
logarithmic from 3 dB above the noise to 5 dB below the maximum
(from Hill et al. 2003).{\it (for higher resolution, 
see paperII-f2.gif)}} \label{fig:dynamspect}\end{figure}

We are grateful to A. Brandenburg,
D. Lambert, F. Matteucci, S. Oey, and G. Tenorio-Tagle for helpful
comments on Section 2; E. Falgarone, D. Hollenbach and J. Le
Bourlot for helpful comments on Section 3; B. Chandran, R.
Jokipii, A. Lazarian, V.S. Ptuskin, and R. Schlickeiser for
helpful comments on Section 4; B. Rickett, S. Spangler, and D.
Stinebring for helpful comments on Section 5 and to A. Hill for
Figure \ref{fig:dynamspect}.

\end{document}